\documentclass[twocolumn,trackchanges]{aastex631}
\usepackage{amsmath}
\usepackage{enumitem}

\shorttitle{Assessing Biosignatures on TOI-700\,d under UV and Pressure Constraints}
\shortauthors{Sumida et al.}
\graphicspath{{./}{figures/}}
\begin{document}

\title{Evaluating Habitability and Biosignature Detection on TOI-700\,d: The Role of UV Environment and Atmospheric Pressure}

\author{Viktor Yuri Doná Sumida}
\affiliation{Center for Radio Astronomy and Astrophysics Mackenzie,
Mackenzie Presbyterian University, \\
Rua da Consolação 930, São Paulo, SP, Brazil}
\affiliation{Jet Propulsion Laboratory, California Institute of Technology \\
4800 Oak Grove Dr, CA, USA}

\author{Raissa Estrela}
% \affiliation{Center for Radio Astronomy and Astrophysics Mackenzie,
% Mackenzie Presbyterian University, \\
% Rua da Consolação 930, São Paulo, SP, Brazil}
\affiliation{Jet Propulsion Laboratory, California Institute of Technology \\
4800 Oak Grove Dr, CA, USA}

\author{Adriana Valio}
\affiliation{Center for Radio Astronomy and Astrophysics Mackenzie,
Mackenzie Presbyterian University, \\
Rua da Consolação 930, São Paulo, SP, Brazil}




\begin{abstract}

\noindent M dwarfs have long been prime targets in the search for habitable exoplanets, owing to their abundance in the galaxy and the relative ease of detecting Earth-sized worlds within their narrower habitable zones. Yet, these low-mass stars can emit high-energy radiation that may gradually erode planetary atmospheres, raising concerns about long-term habitability. TOI-700, a relatively quiescent M dwarf that hosts four known planets, stands out due to its Earth-sized TOI-700\,d in the star’s habitable zone.
Here, we assess whether a habitable environment can be sustained on TOI-700\,d by analyzing different UV flux levels and atmospheric pressures. We focus on two atmospheric scenarios -- one analogous to the Archean Earth and another representing a modern Earth-like environment -- using a 1D photochemistry-climate model. Our results indicate that all simulated cases can maintain temperatures compatible with liquid water on the surface. However, the dominant photochemical pathways differ substantially with UV levels: under low-UV conditions, haze formation in the Archean-like atmosphere provides the main UV shielding, whereas under intensified UV, ozone production in the modern-like atmospheres can protect the surface from harmful doses. Interestingly, although haze can impede the detection of certain biosignatures, such as CH$_4$, CO$_2$ and O$_2$, it also enhances the overall atmospheric signal by increasing scattering and transit depth, potentially aiding in revealing the presence of an atmosphere. These findings underscore the dual role of hazes as both a challenge for biosignature detection and a potential protection of surface habitability.

\end{abstract}

%\keywords{Extrasolar rocky planets (511) --- Exoplanet atmospheres (487) --- Habitable planets (695) --- Ultraviolet astronomy (1736) --- Photodissociation reactions (2266)}


\section{Introduction}

In recent years, red dwarf stars have emerged as primary targets in the search for extraterrestrial life due to their extended lifespans. 
Although the extended pre–main-sequence contraction phase of M dwarf stars may desiccate or erode the atmospheres of planets within their habitable zones, potentially triggering a runaway greenhouse and compromising early habitability \citep{Luger.and.Barnes.2015AsBio..15..119L}, these effects occur mainly during the early stages of stellar evolution. As M dwarfs age, their activity declines substantially, which could allow secondary atmospheres or late-formed planets to sustain habitable environments \citep{Gale2017}.

The Transiting Exoplanet Survey Satellite \citep[TESS,][]{Ricker2015} has played a key role in identifying terrestrial-sized planets, particularly those in the habitable zones of M-dwarfs, which are prime candidates for atmospheric characterization due to their favorable star-to-planet size ratios and close-in habitable zones that enhance transit detectability \citep{Shields16}. These characteristics make planets orbiting M-dwarfs strong candidates for biosignature searches with instruments such as the James Webb Space Telescope (JWST) and future Extremely Large Telescopes (ELTs).

Despite these advantages, characterizing exoplanet atmospheres remains challenging, requiring more detailed observational and theoretical studies. Key biosignature gases -- including O$_2$, CH$_4$, H$_2$O, N$_2$O, O$_3$, and CO$_2$ -- are fundamental for climate regulation, protection from harmful radiation, and the potential indication of biological activity \citep{Airapetian2017}. However, ultraviolet (UV) radiation from M-dwarfs, particularly in the UVB (280--315\,nm) and UVC (100--280\,nm) ranges, can significantly affect atmospheric chemistry and the habitability of planets in close orbits.

The TOI-700 system, an M2.5V dwarf star, hosts four known exoplanets, two of which -- TOI-700\,d and TOI-700\,e -- reside in its habitable zone \citep{Rodriguez2020, Gilbert.et.al.2023ApJ...944L..35G}. TOI-700\,d, a potentially Earth-sized planet ($\sim1.14\,R_\oplus$), receives about 86\% of the Earth's insolation and is of particular interest for habitability studies. Although planets around M-dwarfs are susceptible to atmospheric erosion due to intense X-ray and extreme ultraviolet (XUV) radiation \citep[e.g.,][]{Nishioka.et.al.2023JGRA..12831405N}, TOI-700 itself appears to be relatively quiescent, with X-ray luminosities comparable to those of the modern Sun \citep{Gilbert2020}. This makes TOI-700\,d an excellent candidate for studying the atmospheric processes that influence planetary habitability.

To evaluate TOI-700\,d's potential habitability, we employed the 1D Photochemical Model coupled to the Atmos Climate model \citep{Arney.et.al.2016AsBio..16..873A}, analyzing both Archean and modern Earth-like atmospheres -- biogenically active scenarios that assume the presence of life -- across pressures of 0.5, 1.0, 2.0, and 4.0 bar. Despite TOI-700 being relatively quiescent during the observed period so far, we simulated two scenarios: one assuming that TOI-700 remains in a quiet state ($f_\mathrm{UV}$=1) and another representing a flare event causing a 10-fold increase in NUV flux ($f_\mathrm{UV}$=10), as estimated by prior studies on M-type stars \citep{Rekhi.et.al.2023ApJ...955...24R}. 

Our work addresses critical but underexplored aspects such as how photochemistry influences atmospheric composition, the potential for false-positive biosignatures, and climate effects under varying levels of UV flux and pressure. These challenges align with the call for advanced atmospheric models by the 2020 Astronomy and Astrophysics decadal survey\footnote{\url{https://www.nationalacademies.org/our-work/decadal-survey-on-astronomy-and-astrophysics-2020-astro2020}} to evaluate star-planet interactions, haze formation, atmospheric dynamics, and escape processes. 
Although detecting TOI-700\,d’s transmission features would require untenably long JWST integration times, the significance of this work is reinforced by the growing focus of observational programs utilizing the James Webb and Hubble Space Telescopes, which aim to probe planetary atmospheres and stellar activity around M dwarfs, such as the Rocky Worlds DDT program\footnote{\url{https://www.stsci.edu/contents/news/jwst/2024/rocky-worlds-ddt-selects-its-first-targets}}.

%A key aspect of our analysis involves evaluating whether TOI-700\,d could sustain life by assessing the ability of its atmosphere to shield against UV radiation and its implications for habitability, which we discuss in \autoref{sec:bio_effect}. Additionally, we investigate the impact of photochemically produced haze on the planet’s atmospheric transmission spectra, since haze can obscure key spectral features of biosignature gases, which we explore in \autoref{sec:transmission_spectra}.

In this study, we assess the habitability potential of TOI-700 d by simulating different atmospheric compositions and evaluating their response to varying stellar UV flux. In Section~\ref{sec:methods}, we outline the methods employed, including the 1D Photochemical Model coupled to the Atmos Climate model used to simulate Archean and modern Earth-like atmospheres under different pressure conditions. Section~\ref{sec:results} presents our results and discussion, including the atmospheric abundance profiles, photochemical pathways, and the impact of UV radiation on surface habitability. A key aspect of our analysis involves evaluating whether TOI-700\,d could sustain life by assessing the ability of its atmosphere to shield against UV radiation and its implications for habitability, which we discuss in Subsection~\ref{sec:bio_effect}. 
Additionally, we investigate the impact of photochemically produced haze on the planet’s atmospheric transmission spectra, since haze can obscure key spectral features of biosignature gases, which we explore in Subsection~\ref{sec:transmission_spectra}. Finally, Section~\ref{sec:conclusion} summarizes our key findings, discusses the broader implications for exoplanetary habitability studies, and highlights future observational prospects.

\section{Methods} \label{sec:methods}

Possible atmospheres for the exoplanet TOI-700\,d were simulated with the 1D Photochemical Model coupled to the Atmos Climate model \citep{Arney.et.al.2016AsBio..16..873A}. The code generates atmospheres with well-defined chemical and climate profiles. 
In the modern Earth-like atmosphere photochemical model, the code incorporates 310 chemical reactions and 72 chemical species, including 11 short-lived species. On the other hand, for the Archean atmosphere, the model incorporates 405 chemical reactions and 77 chemical species, with 9 of them being short-lived. 

The initial atmospheric state is established by initially executing the photochemical model, which relies on user-specified boundary conditions. After the photochemical model converges, its output files containing altitude, pressure, gas mixing ratios, haze particle sizes, and haze number densities are transferred to the climate model. The climate model utilizes the photochemical model’s solution as its starting point and continues running until it also achieves convergence. Subsequently, the updated temperature and water vapor profiles are fed back into the photochemical model. This iterative process continues until convergence is attained. Detailed descriptions of the model can be found in \cite{Arney.et.al.2016AsBio..16..873A}, and are publicly accessible\footnote{\url{https://github.com/VirtualPlanetaryLaboratory/atmos}}.

To accurately simulate planetary environments and observable properties, the models require comprehensive data on both planetary and stellar characteristics. This includes detailed information on stellar parameters and spectra, as well as planetary physical and orbital parameters. Additionally, environmental data are crucial for refining the simulations. 
Here, we used the  stellar and planetary parameters reported by \cite{Gilbert2020}. 
For the stellar spectrum, Atmos provides twenty-one star spectra encompassing a range of spectral types. Given that TOI-700 is classified as an M2.5V star, the closest spectral type available is M3.0V, which corresponds to GJ\,581 \citep{Turnbull.2015arXiv151001731T}. 

Following \cite{Arney.et.al.2016AsBio..16..873A} and \cite{Meadows.et.al.2018}, organic haze particles were considered to possess a fractal shape in both photochemical and climate models, to the detriment of spherical (nonfractal or classical Mie) particles. 
The solar zenith angles (SZAs) used in these models were selected to realistically represent the global averaged insolation. In particular, 60$^\circ$ in the climate model corresponds to global mean insolation (accounting for S/4 and diurnal effects), while 45$^\circ$ in the photochemical model -- based on \cite{Segura.et.al.2003AsBio...3..689S} -- reproduces the modern Earth's ozone column given the employed chemistry. These values are common approximations, although more sophisticated methods (e.g., Gaussian integration) could improve insolation estimates if full ozone photochemistry were implemented.

Another key parameter of Atmos is the UV amplification factor ($f_\mathrm{UV}$). This parameter plays an important role in simulating stellar flares, as it boosts the UV flux reaching the top of the atmosphere. In simple terms, it determines the relationship between $F_\mathrm{flare,UV}$ and $F_\mathrm{q,UV}$, where $F_\mathrm{q}$ represents the quiescent UV radiation flux. \cite{Rekhi.et.al.2023ApJ...955...24R} provided estimates for the median  top-of-atmosphere (TOA) NUV irradiances for planets located in the habitable zones of M0 to M6 dwarfs using the GALEX sample. They also determined threshold flare amplitudes and their corresponding frequencies necessary to achieve the TOA NUV irradiance levels similar to those of the young Sun. According to these authors, the exoplanetary NUV irradiance, accounting for flares, is within one order of magnitude of the young Earth's NUV irradiance for host stars post-M2. Flares reaching the young Earth's irradiance level occur multiple times a day for stars later than M3.
According to Table 3 in the study by \cite{Rekhi.et.al.2023ApJ...955...24R}, the ratio of flare to quiescent NUV flux ($F_\mathrm{flare,UV}/F_\mathrm{q,UV}$) is 12.4 for an M2 star and 7.9 for an M3 star. Since TOI-700 is classified as an M2.5 star, we extrapolated an intermediate value of approximately 10 for $f_\mathrm{UV}$ in our simulations.

Although our Atmos simulations for various atmospheres of TOI-700\,d  provide valuable insights, it's imperative to determine the strength of potentially observable spectral features. To accomplish this, we use the calculated values of gas mixing ratios of H$_2$O, CH$_4$, C$_2$H$_6$, CO$_2$, O$_2$, O$_3$, CO, H$_2$CO, HNO$_3$, NO$_2$, SO$_2$, N$_2$O and N$_2$, and aerosol parameters derived from the Atmos simulations. These inputs are fed into the Planetary Spectrum Generator \citep[PSG,][]{Villanueva.et.al.2018JQSRT.217...86V} module known as GlobES\footnote{\url{https://psg.gsfc.nasa.gov/apps/globes.php}} (Global Exoplanet Spectra), which is designed for synthesizing both transmission and emission spectra. The PSG integrates advanced radiative transfer techniques to accurately generate comprehensive synthetic spectra 
%to further our understanding of the atmospheric properties 
of TOI-700\,d and to assess the feasibility of observing spectral features in its atmosphere.

\section{Results and discussion} \label{sec:results}

In Subsection~\ref{sec:photoclima}, we present the results and discuss the implications of the atmospheric abundance profiles of the models, along with their respective transmission spectra. Then in Subsection~\ref{sec:bio_effect}, we  assess the habitability potential of TOI-700\,d by conducting an analysis of the impact of UV radiation on the planet's surface. To complete the analysis, in Subsection~\ref{sec:transmission_spectra}, we present the simulated transmission spectra derived from our models and discuss their observational implications.

\subsection{Photochemical pathways in different conditions} \label{sec:photoclima}

\begin{figure*}
    \centering       
    \includegraphics[width=0.3\textwidth]{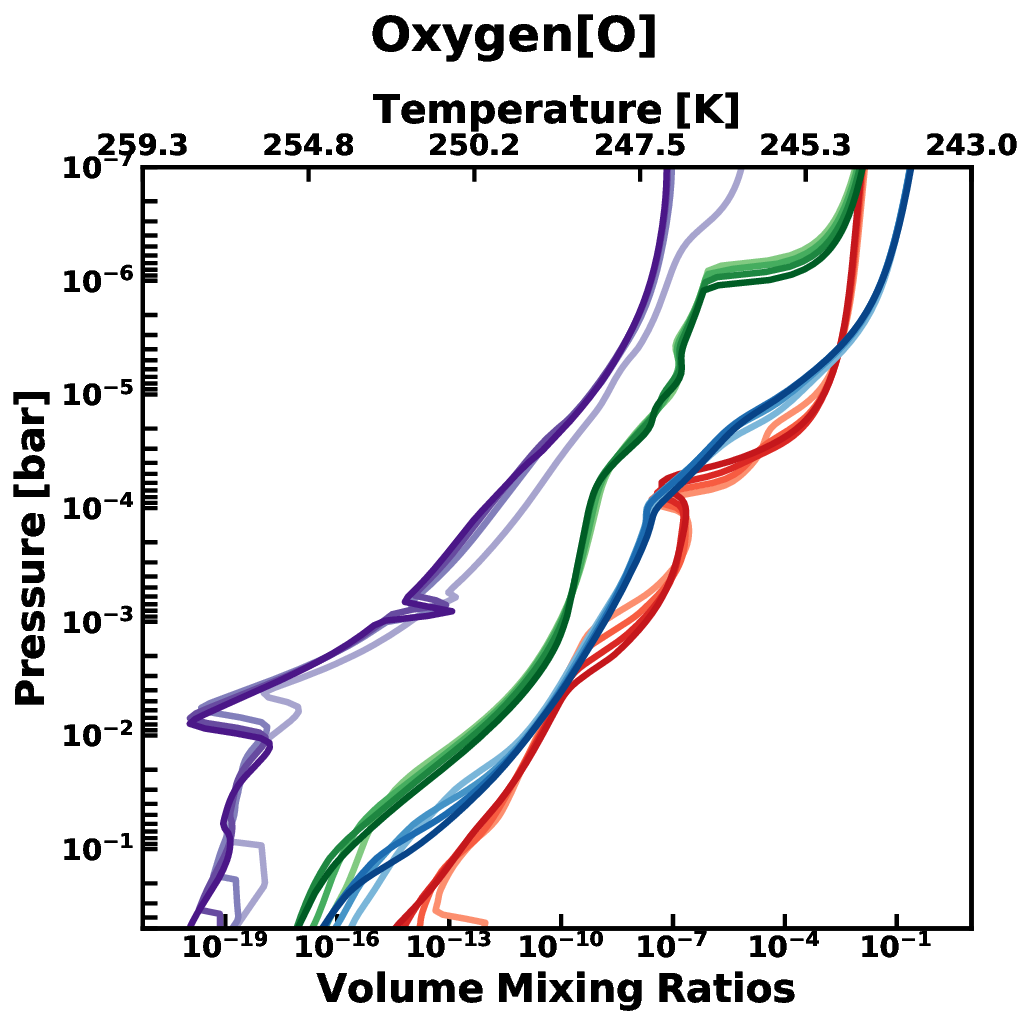}
    \includegraphics[width=0.3\textwidth]{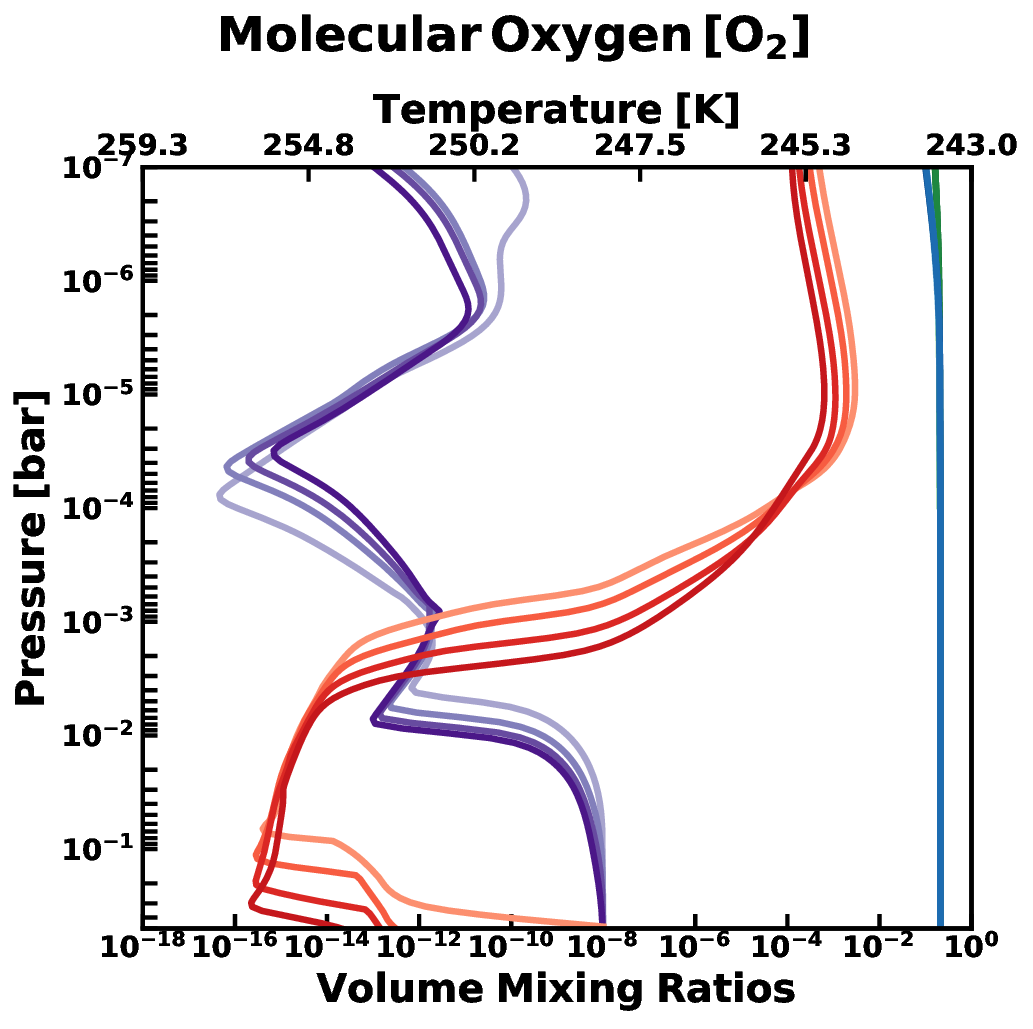}
    \includegraphics[width=0.3\textwidth]{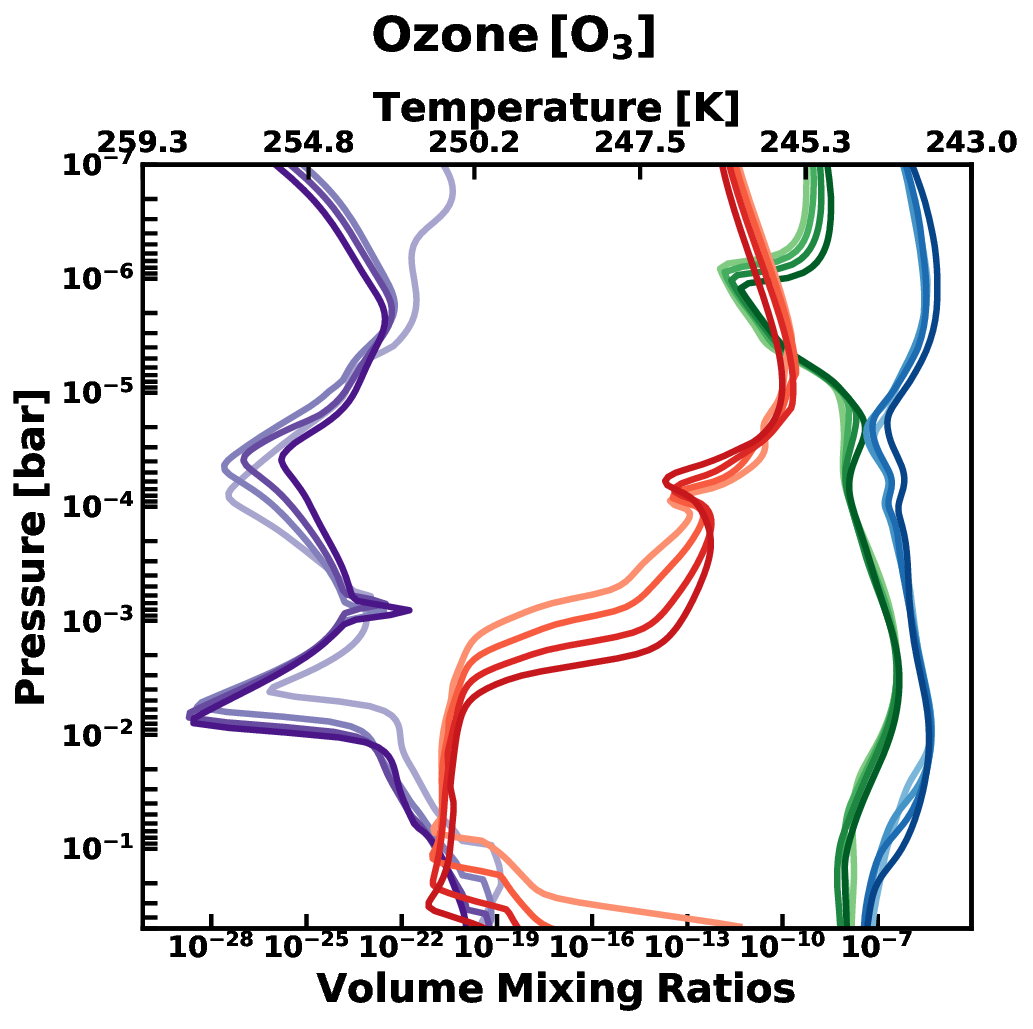}
    \includegraphics[width=0.3\textwidth]{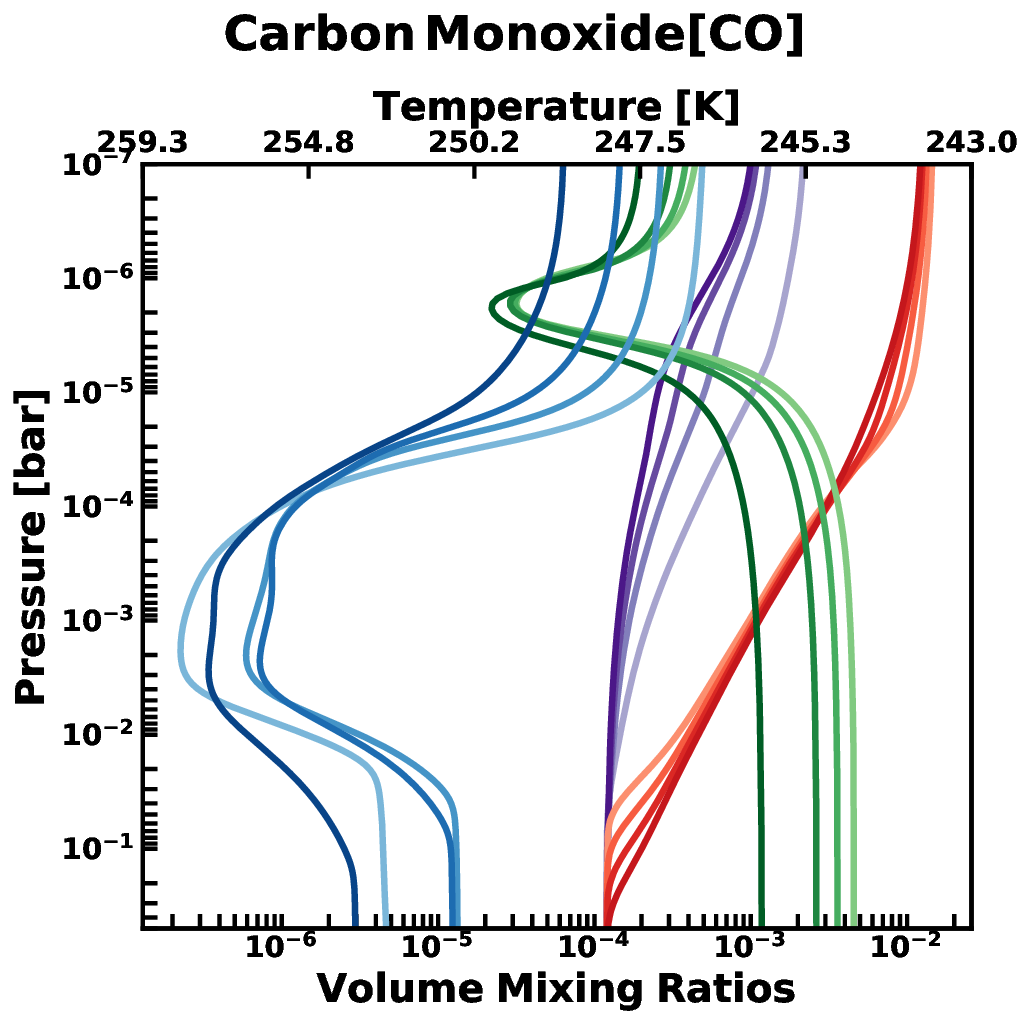}
    \includegraphics[width=0.3\textwidth]{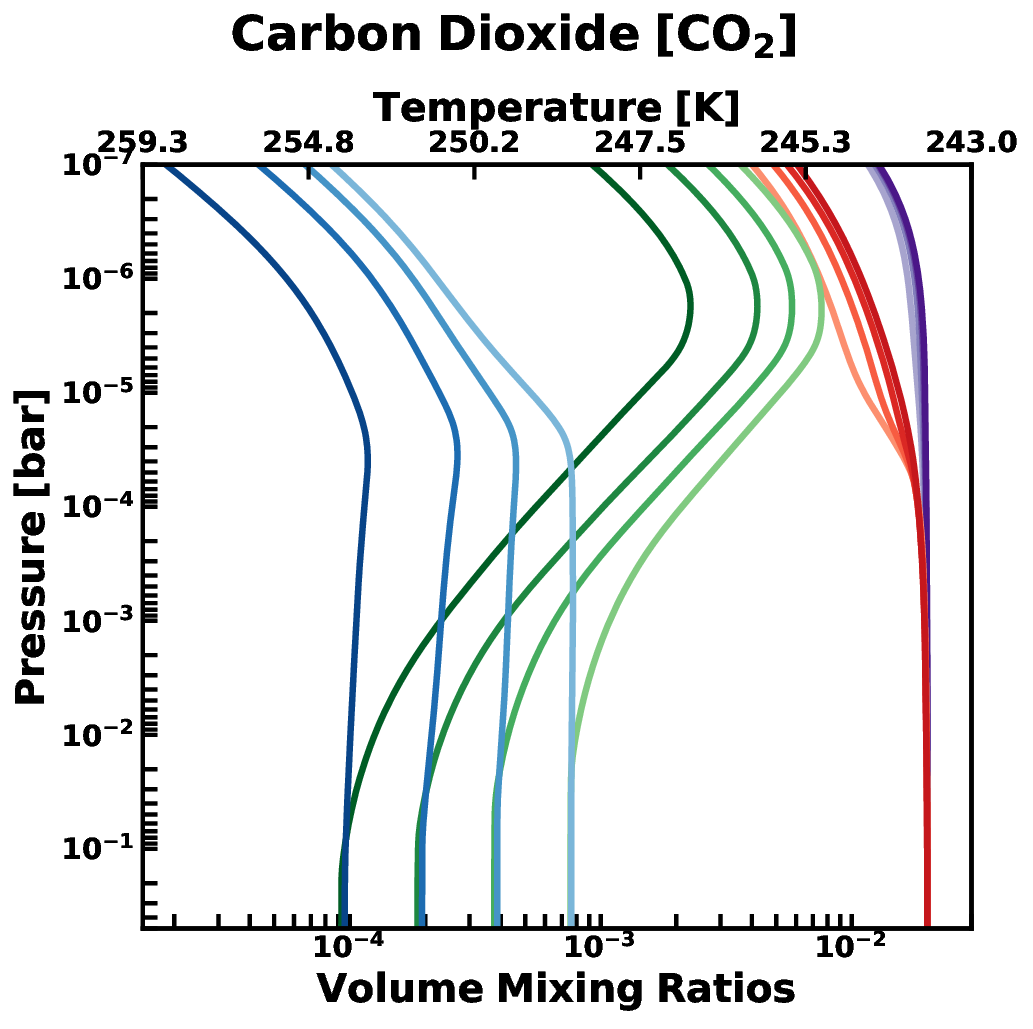}
    \includegraphics[width=0.3\textwidth]{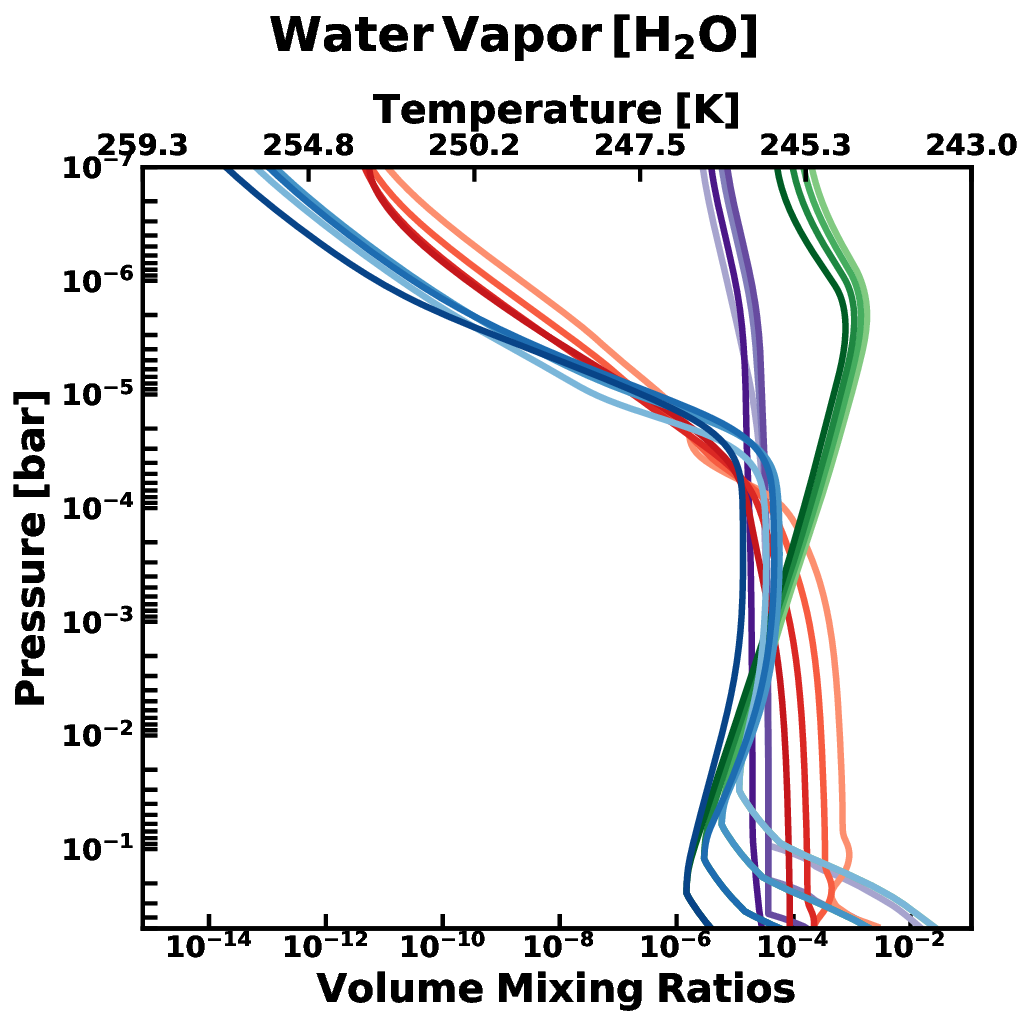}
    \includegraphics[width=0.3\textwidth]{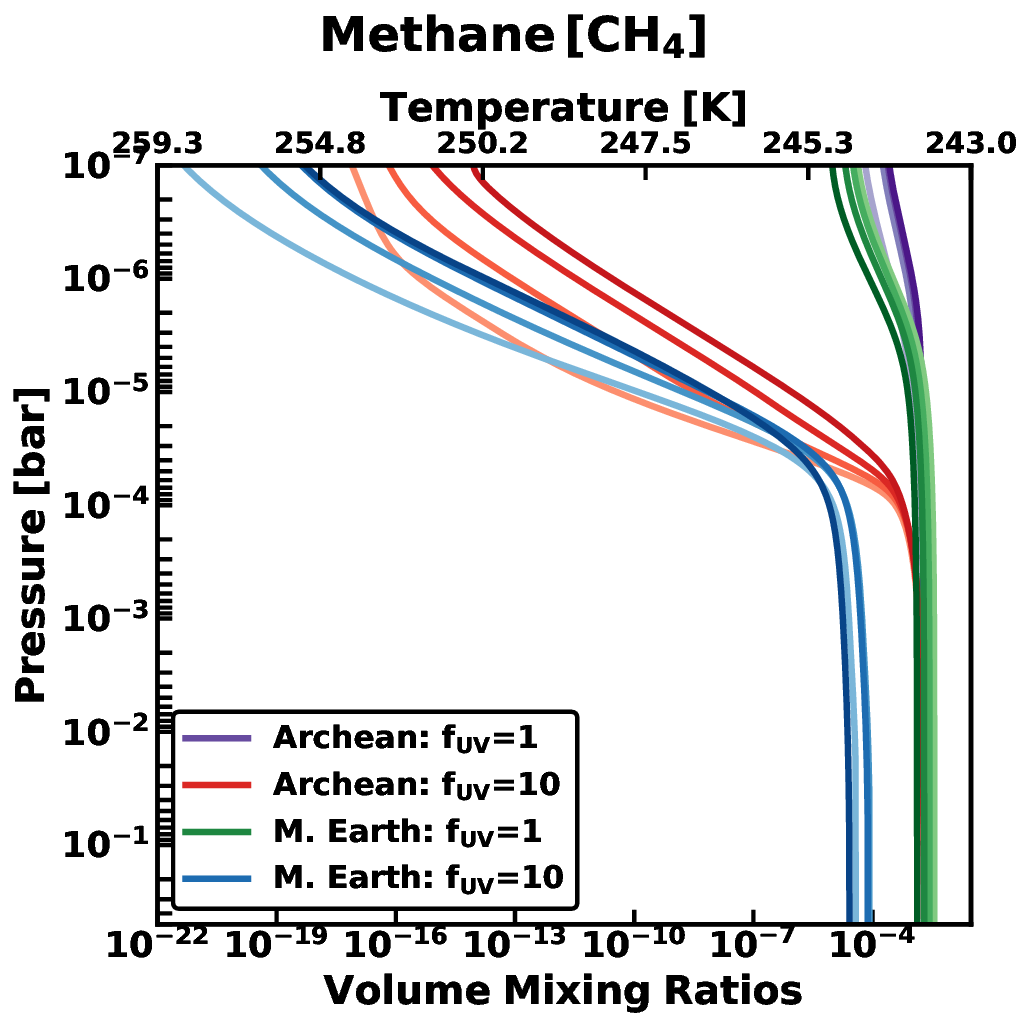}
    \includegraphics[width=0.3\textwidth]{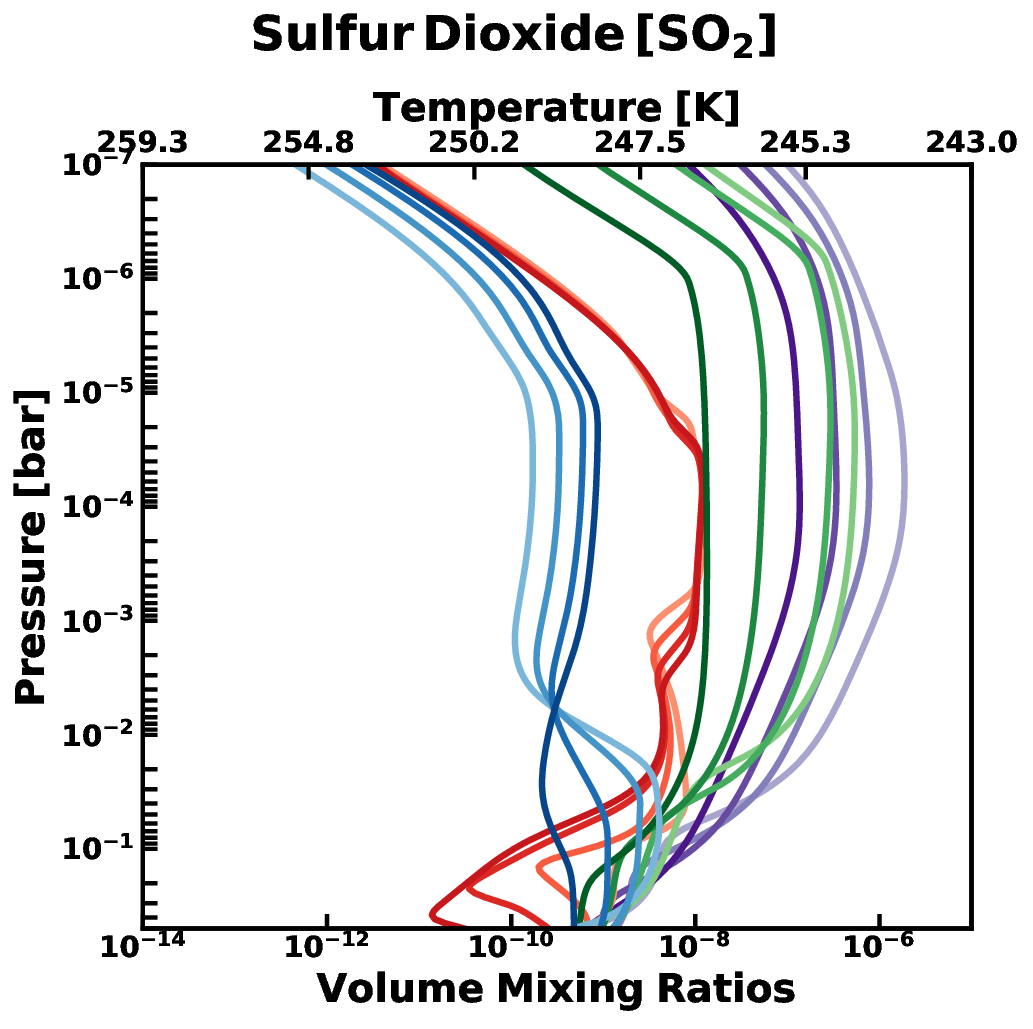}
    \includegraphics[width=0.3\textwidth]{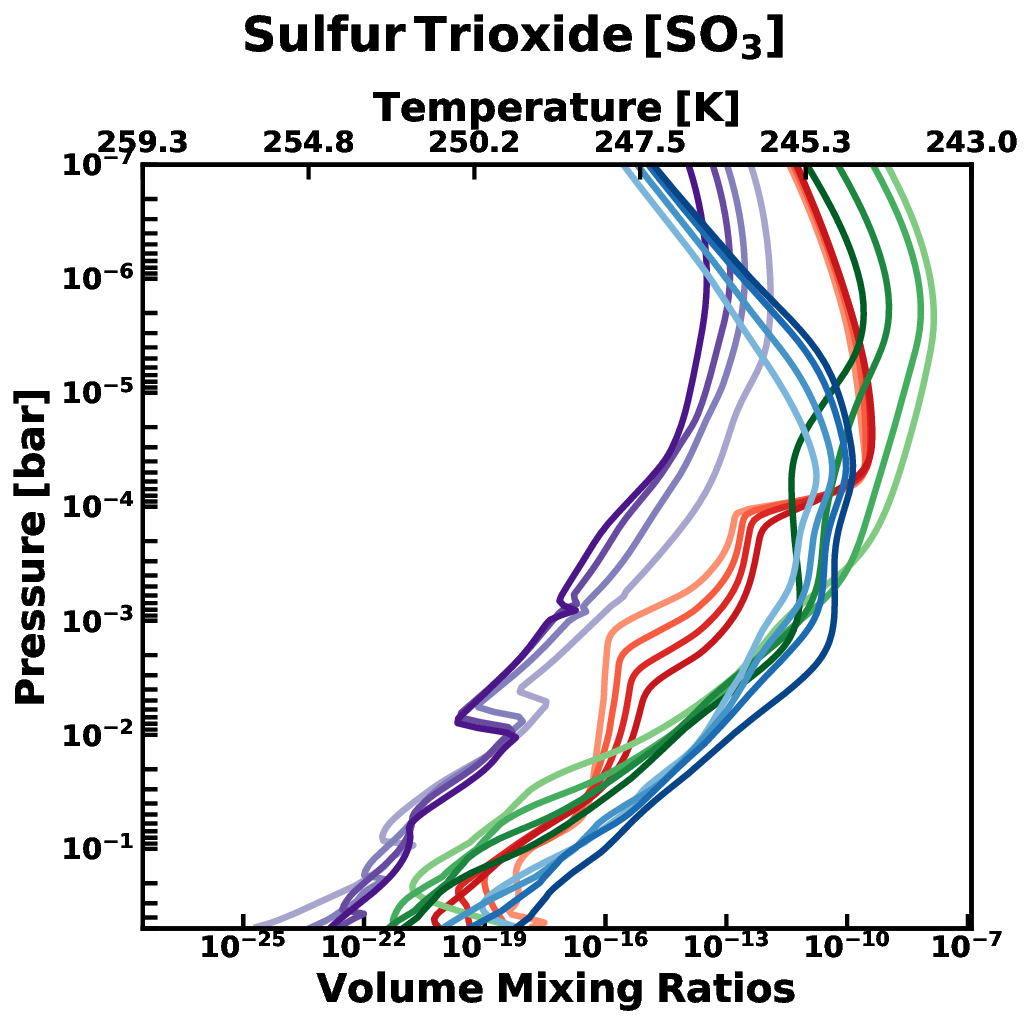}
    \caption{Each panel, from the upper left to the lower right, represents the atmospheric pressure  of O, O$_2$, O$_3$, CO, CO$_2$, H$_2$O, CH$_4$, SO$_2$ and SO$_3$ as a function of mixing ratios. 
    The initial conditions and the produced surface mixing ratios can be seen either in \autoref{tab:inital_produced_mix_ratios} or in the files available at DOI: \href{https://doi.org/10.5281/zenodo.13947863}{10.5281/zenodo.13947863}.
    The Archean atmospheres with $f_\mathrm{UV}$=1 and $f_\mathrm{UV}$=10 are depicted in purple and red, while modern Earth-like atmospheres with $f_\mathrm{UV}$=1 and $f_\mathrm{UV}$=10 are shown in green and blue, respectively. The different surface pressures, i.e., 4.0, 2.0, 1.0, and 0.5\,bar, are represented by tones gradually shifting towards lighter shades, indicating a transition from higher to lower pressure.}
    \label{fig:abundance_profiles}
\end{figure*}

Profiles of possible atmospheres for TOI-700\,d are simulated, based on both Archean and modern Earth-like atmospheres. Our simulations incorporate variations in atmospheric pressure at 0.5, 1.0, 2.0, and 4.0\,bar, and compare two distinct scenarios: one with $f_\mathrm{UV}=1$ and the other with $f_\mathrm{UV}=10$. The resulting abundance profiles for key molecules of interest are presented in \autoref{fig:abundance_profiles}. 
All figures presented here adhere to the same color scheme for the atmospheres: Archean and modern Earth-like with $f_\mathrm{UV}$=1 are represented in purple and green, while Archean and modern Earth-like atmospheres with $f_\mathrm{UV}$=10 are depicted in red and blue, respectively. The varying pressures, namely 4.0, 2.0, 1.0, and 0.5\,bar, are indicated by a gradual transition towards lighter shades, illustrating a shift from higher to lower pressure. 
%The abundance profiles obtained for some molecules of interest are displayed in \autoref{fig:abundance_profiles}.

The choice of methane to carbon dioxide ratios in atmospheric models reflects the conditions from different geological periods and their relevance to climate processes. 
For Archean Earth scenarios, a CH$_4$/CO$_2$ ratio above 0.1 is widely used as a criterion since it enables significant organic haze formation in primitive atmospheres \citep[e.g.,][]{Haqq-Misra.et.al.2008AsBio...8.1127H, Arney.et.al.2017ApJ...836...49A, Meadows.et.al.2018, Mak.et.al.2023JGRD..12839343M}.

For TOI-700\,d, our Archean atmospheric model adopts a CH$_4$/CO$_2$ ratio of approximately 0.5, a value chosen to reflect the enhanced methane accumulation expected around M-dwarf stars. This selection is based on the findings of \cite{Meadows.et.al.2018}, who demonstrated that a CH$_4$/CO$_2$ ratio of approximately 0.3 is necessary to trigger the formation of organic haze on Proxima Centauri\,b. 
In an oxygen-poor, Archean-like atmosphere, the primary pathways for CH$_4$ destruction involve OH radicals formed via H$_2$O photolysis and O atoms generated by CO$_2$ photolysis.
However, reduced UV emission around 300\,nm typical of M-dwarf stars leads to a diminished photolysis rate, allowing methane to accumulate. In our model for TOI-700\,d, as methane becomes more dominant, aerosol production intensifies, leading to a thickening haze that significantly reduces the penetration of starlight into the planet's surface, resulting in substantial cooling \citep[see][]{Haqq-Misra.et.al.2008AsBio...8.1127H}.
Note that this Archean-like composition is included solely as a limiting case, allowing us to probe the climatic and UV-shielding consequences of a methane-rich, low-oxygen atmosphere; sustaining CH$_4$/CO$_2 \approx 0.5$ would in reality require exceptionally vigorous biogenic activity or intense geological outgassing, so we treat it as an exploratory end-member rather than a prediction of TOI-700 d’s actual state.

For the modern-Earth analog, we adopt recent near-surface mixing ratios (CH$_4$ $\approx$ 1.8 ppmv; CO$_2$ $\approx$ 385 ppmv), corresponding to CH$_4$/CO$_2$ $\approx$ 4.7$\times$10$^{-3}$, as an empirical observational baseline for comparison \citep[e.g.,][]{IPCC.2021AR6}\footnote{Over geologic timescales, atmospheric CO$_2$ on Earth is regulated by the carbonate--silicate feedback and has been further perturbed in the industrial era; CH$_4$, in contrast, is primarily controlled by biogenic sources and removal by atmospheric oxidants. In this work we use the observed modern-Earth CH$_4$/CO$_2$ ratio only as an empirical baseline for comparison; our conclusions are insensitive to modest variations within the explored range (10$^{-3}$--10$^{-2}$).}.
Furthermore, we specified the lower boundary conditions of the photochemical model to reproduce observed surface-level concentrations of biogenic trace gases on modern Earth, for example, N$_2$O around 0.3\,ppmv in the lower atmosphere. This value corresponds to estimated modern biogenic fluxes that sustain such concentrations. In this way, the modern Earth scenario in our model begins with N$_2$O levels aligned with present-day atmospheric observations \citep[see][]{IPCC.2021AR6}.
In such an atmosphere, like that of modern Earth, methane undergoes rapid oxidation through reactions with hydroxyl radicals (OH):
\begin{equation}
\begin{gathered}
\text{O}_2 + h\nu \; (\lambda < 200\,\textrm{nm})  \xrightarrow{} :\text{O}\,+ \text{O} \\
:\text{O} + \text{O}_2     \xrightarrow{} \text{O}_3 \\
\text{O}_3 + h\nu \; (\lambda < 310\,\textrm{nm})  \xrightarrow{} \text{O}_2 + \text{O}^* \\
\text{O}^* + \text{H}_2\text{O}    \xrightarrow{} 2\;\text{OH} \\
\text{CH}_4 + \text{OH}    \xrightarrow{} \text{CH}_3 + \text{H}_2\text{O}
\end{gathered}
\label{eq:CH4}
\end{equation}
This lower CH$_4$/CO$_2$ ratio is more suitable for current Earth-like conditions, where oxygen significantly reduces methane's presence, limiting its capacity to contribute to haze formation. By modeling both Archean and modern conditions, we can observe the shift in haze dynamics and its impact on the planet's climate system.

%\textbf{\textcolor{red}{VS: Nao sei se e valido colocar este paragrafo:} Similarly to methane, the distinctive atmospheric conditions of planets orbiting M-dwarf stars can lead to significantly higher concentrations of gases such as nitrous oxide (N$_2$O) compared to Earth. Although elevated N$_2$O levels have been considered a robust biosignature on Earth-like planets orbiting G-type stars -- where its source is nearly entirely biogenic -- the situation for M-dwarf planets is more complex. In such environments, both biogenic and abiotic sources (e.g., energetic particles from the host star actively induce the formation, as \citep{Airapetian.et.al.2016NatGe...9..452A} observed on early Earth.) can contribute to N$_2$O accumulation. In our atmospheric models of TOI-700\,d with a modern Earth-like composition, however, we assume that the enhanced N$_2$O concentration arises primarily from biogenic sources. The concentration of N$_2$O increases substantially in our models, with the degree of enhancement varying considerably with UV radiation levels and atmospheric pressure. At UV levels of $f_\mathrm{UV}=1$, N$_2$O concentrations are 32, 26, 17, and 10 times higher than Earth's for pressures of 0.5, 1, 2, and 4\,bar, respectively. When UV levels are ten times higher ($f_\mathrm{UV}=10$) due to flaring, the increase is more moderate, with N$_2$O concentrations rising by factors of 5.6, 5.1, 4.7, and 4.0 for the same pressures. These results highlight that, in low-UV environments, N$_2$O can accumulate significantly.}

\autoref{tab:inital_produced_mix_ratios} in the appendix shows the initial conditions and produced surface mixing ratios of the nine major long-lived species for the Archean and modern Earth-like atmosphere models, at a pressure of 1\,bar. For more detailed analysis, including simulations with varying pressures and UV fluxes, the complete data set is available at DOI: \href{https://doi.org/10.5281/zenodo.13947863}{10.5281/zenodo.13947863}

The exact composition of the Archean atmosphere when life first emerged on Earth remains a subject of ongoing debate. Nevertheless, the interplay between the atmosphere and geological cycles has left discernible traces that help identify the main atmospheric gases of this era. Combined geological and atmospheric modeling suggests that the Archean atmosphere (4--2.5\,Gyr ago) may have contained approximately 10 to 2500 times the modern amount of CO$_2$ and 100 to 10$^4$ times the modern amount of CH$_4$ \citep[see][]{Catling20}.
Although the plausibility of extremely high methane levels in prebiotic eras remains debated, recent studies point to potential abiotic mechanisms. \cite{Wogan.et.al.2023PSJ.....4..169W}, for instance, argue that major impact events could substantially increase CH$_4$ concentrations, allowing for episodic surges in atmospheric methane even in the absence of biological activity.

As shown in the bottom-left panel of \autoref{fig:abundance_profiles}, the methane concentration in the two model atmospheres with $f_\mathrm{UV}=1$ and in the Archean atmosphere with $f_\mathrm{UV}=10$ is roughly 10$^3$ times higher than in the modern Earth's atmosphere with $f_\mathrm{UV}=10$. Notably, both the Archean and modern Earth models with $f_\mathrm{UV}=10$ exhibit an exponential decrease in methane abundance with increasing altitude. In atmospheres exposed to elevated UV flux ($f_\mathrm{UV}=10$), methane undergoes a more pronounced decline in its mixing ratio owing to enhanced CH$_4$ oxidation by UV radiation. By contrast, in simulations with normal UV flux ($f_\mathrm{UV}=1$), the mixing ratio of methane still decreases with altitude, albeit more gradually, reflecting a comparatively lower oxidation rate. These differing rates of CH$_4$ oxidation under varying UV flux conditions lead to distinct vertical mixing ratio profiles across the modeled atmospheres.

\begin{figure*}
    \centering       
    \includegraphics[width=0.99\textwidth]{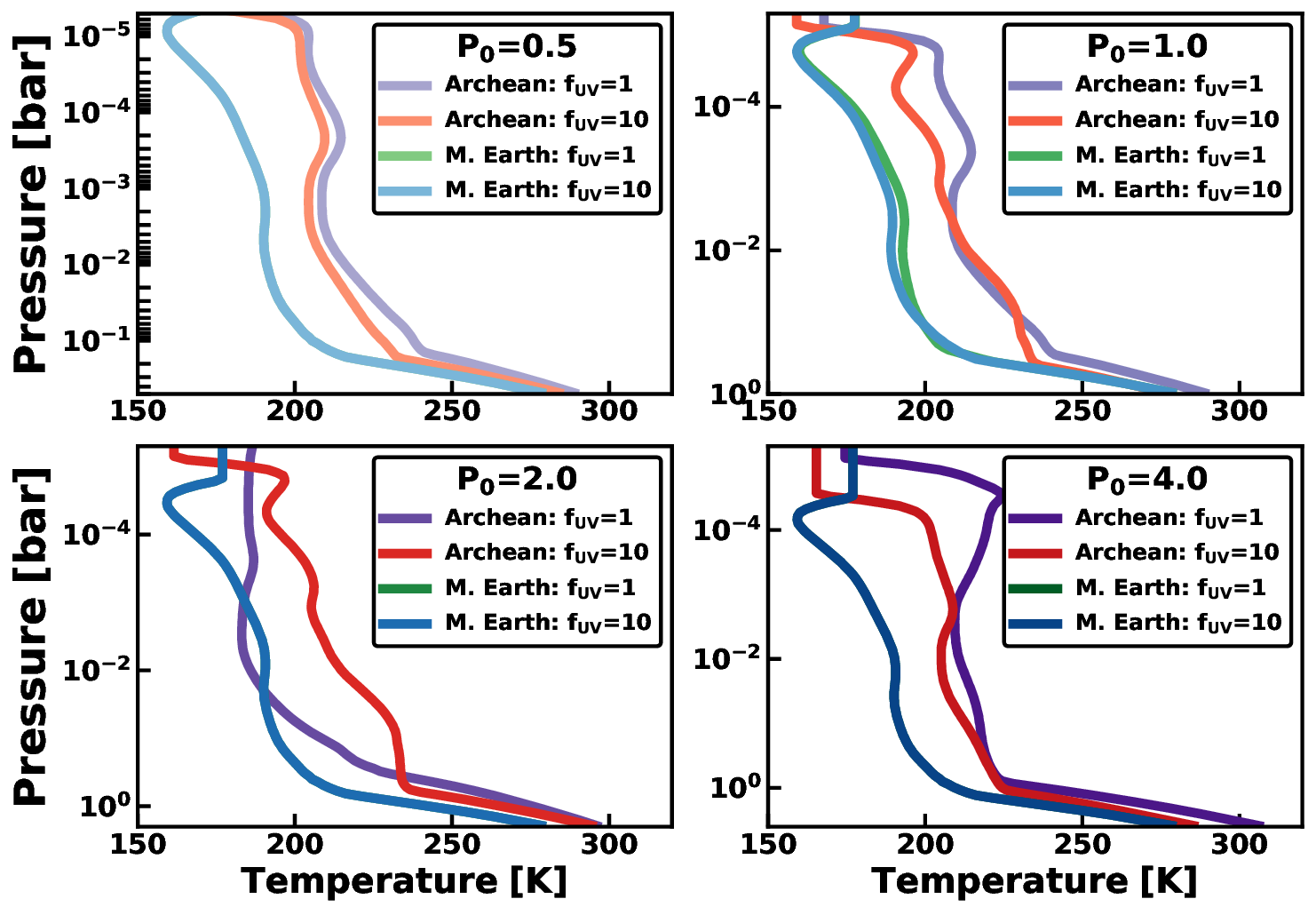}
    \caption{Atmospheric pressure profiles plotted against temperature for different pressures. The Archean atmospheres with $f_\mathrm{UV}$=1 and $f_\mathrm{UV}$=10 are depicted in purple and red, while modern Earth-like atmospheres with $f_\mathrm{UV}$=1 and $f_\mathrm{UV}$=10 are shown in green and blue, respectively. The Modern Earth atmosphere models of $f_\mathrm{UV}$=1 and $f_\mathrm{UV}$=10 can only be distinguished from one another in the case where the surface pressure ($P_0$) equals 1 (top right panel), otherwise for all other pressures, the curve for $f_\mathrm{UV}$=10 (blue) superimposes on the $f_\mathrm{UV}$=1 (green) curve.}
    \label{fig:atmospheric_params}
\end{figure*}

In our Archean--like scenarios, the surface temperature is higher than in the modern Earth case (see \autoref{fig:atmospheric_params}). Although CH$_4$ is more radiatively efficient per molecule than CO$_2$, primarily at trace-level concentrations typical of present-day Earth, the total greenhouse forcing in the Archean simulations is dominated by the substantially higher absolute abundances of both CH$_4$ and CO$_2$ \citep[see][]{Kiehl.and.Dickinson.1987JGR....92.2991K, Etminan.et.al.2016GeoRL..4312614E}. This enhanced greenhouse effect more than offsets the reduced stellar energy input to TOI-700\,d, allowing surface temperatures that are comparable to, or even exceed, those required for the stability of liquid water.
However, without continuous replenishment these gases cannot sustain such warming over geological timescales: CH$_4$ is photochemically oxidized to CO and CO$_2$, while CO$_2$ is gradually sequestered by carbonate--silicate weathering and ocean uptake.

Consequently, geological activity on a planet is necessary to balance these gases through the carbonate-silicate cycle, which involves volcanic activity, plate tectonics, and erosion. The lack of greenhouse gas replacement by geological activity has contributed to Mars' current temperatures of -50$^\circ$\,C on its surface and a pressure of only 1\% of Earth's atmospheric pressure today \citep{Segura-Navarro2005}. 
Regarding temperature under haze conditions, cooling is minimized around M-dwarfs. These stars emit energy primarily at wavelengths where organic hazes are relatively transparent \citep{Arney.et.al.2017ApJ...836...49A}.

In all scenarios, the estimated temperatures suggest conditions conducive to liquid water on the surface of TOI-700\,d. The lowest temperature, approximately 278\,K, is associated with Modern Earth-like models with $f_\mathrm{UV}$=1 and $f_\mathrm{UV}$=10, irrespective of surface pressure. Conversely, the highest temperature of 306\,K is observed in the $f_\mathrm{UV}$=1 Archean model with a surface pressure of 4\,bar. Detailed temperature values for each model are provided in \autoref{tab:temp}. 

Studies on Archean Earth suggest that extensive ocean fractions could persist for globally averaged surface temperatures as low as 250–260\,K \citep[e.g.,][]{Wolf2013, Arney.et.al.2016AsBio..16..873A}. However, such studies are based on planets orbiting Sun-like stars. 
Around M dwarfs, intense XUV flux and frequent stellar flares can significantly enhance atmospheric escape, leading to substantial water loss over time. 
While previous studies \citep[e.g.,][]{do.Amaral.et.al.2022ApJ...928...12D} have emphasized the role of stellar flares in driving early atmospheric escape, we also acknowledge the cumulative effects of stellar luminosity evolution. Late M dwarfs such as TOI-700 can brighten significantly over time. According to our XUV flux calculations (see Equation\,1 in Jackson et al. \citeyear{Jackson.et.al.2012MNRAS.422.2024J} and Equation\,22 in Owen and Wu \citeyear{Owen.and.Wu.2017ApJ...847...29O}), TOI-700\,d received an estimated XUV flux of $\sim 574$\,erg\,s$^{-1}$\,cm$^{-2}$ at 100\,Myr\footnote{Stellar X-ray emission typically undergoes an initial saturation phase lasting tens to hundreds of millions of years, followed by a gradual decline. This evolutionary trend is observed across all stellar spectral types \citep{Jackson.et.al.2012MNRAS.422.2024J, Shkolnik.and.Barman.2014AJ....148...64S, McDonald.et.al.2019ApJ...876...22M}.} -- over 15 times higher than its present--day value of $\sim 37$\,erg\,s$^{-1}$\,cm$^{-2}$. This elevated high-energy input during the first few hundred million years of evolution could have driven intense hydrodynamic escape, especially if TOI-700\,d originally possessed a volatile-rich envelope. Our results thus reinforce the scenario raised by \cite{Luger.and.Barnes.2015AsBio..15..119L}, in which secular stellar brightening -- independent of flares -- may lead to substantial or even complete devolatilization of initially habitable-zone planets around M dwarfs.

Similarly to \cite{Meadows.et.al.2018}, attempting to run the climate model to a converged state using the same top-of-atmosphere pressure as the photochemical model proved unfeasible. This limitation stems from instabilities within the model, yet it does not impact the resultant photochemistry, as indicated by \cite{Arney.et.al.2016AsBio..16..873A}. Consequently, when the climate model transfers its temperature and water profiles to the photochemical model, they remain fixed at their values at the top of the climate grid, forming isoprofiles above this grid's upper limit. This effect is illustrated in \autoref{fig:atmospheric_params}, where an increase in surface pressure leads to isoprofiles occurring at higher pressures.

\begin{table}[]
    \caption{Surface Temperature Estimate for TOI-700\,d.}
    \begin{center}
   \begin{tabular}{l|lcc}
   \hline
    \multicolumn{1}{c|}{Pressure} & Atmospheric & $f_\mathrm{UV}$ & Temperature \\
    \multicolumn{1}{c|}{[bar]} & Model & & [K] \\
    \hline
    0.5 & Archean & 1 & 288.5 \\
        & Archean & 10 & 283.9 \\
        & Modern Earth & 1 &  278.4 \\
        & Modern Earth & 10 &  278.4 \\
    \hline
        1.0 & Archean & 1 & 288.8\\
        & Archean & 10 & 279.3 \\
        & Modern Earth & 1 & 278.5 \\
        & Modern Earth & 10 & 278.4 \\
    \hline
        2.0 & Archean & 1 & 296.0 \\
        & Archean & 10 & 294.5 \\
        & Modern Earth & 1 & 278.4 \\
        & Modern Earth & 10 & 278.3 \\
    \hline
        4.0 & Archean & 1 & 306.2 \\
        & Archean & 10 & 285.4 \\
        & Modern Earth & 1 & 278.4 \\
        & Modern Earth & 10 & 278.3 \\
    \hline
    \end{tabular}
    \end{center}
    \label{tab:temp}
\end{table}

An apparently puzzling feature in \autoref{fig:abundance_profiles} is the
behavior of CO in the modern Earth scenario run with
$f_{\mathrm{UV}}=1$.  Below the homopause ($P\!\gtrsim\!10^{-5}$\,bar) the
gas-phase carbon inventory -- $f_{\mathrm{CO_2}}+f_{\mathrm{CO}}+f_{\mathrm{CH_4}}$ -- is conserved in all
scenarios, yet the partitioning among the three species depends on the local UV field and on the oxidative state of the atmosphere:

\begin{enumerate}[label=\roman*]

\item Modern Earth, $f_{\mathrm{UV}}=1$: a modest O$_3$ layer removes most 200--310\,nm photons. The resulting near-UV deficit almost quenches CO$_2$ photolysis above $10^{-6}$\,bar, eliminating the main \emph{in situ} source of CO, and keeps OH production low; the residual OH therefore acts as a net sink via $\mathrm{CO+OH\rightarrow CO_2}$.  With its source suppressed, CO decreases with altitude and the lost carbon reappears as CO$_2$.

\item Archean, $f_{\mathrm{UV}}=1$: in the absence of O$_2$/O$_3$, CO$_2$ photolysis operates throughout the upper column and the CO profile increases.  Removal of carbon into hydrocarbon haze occurs between $10^{-6}$ and $10^{-5}$\,bar, but the fraction sequestered remains $\lesssim10^{-4}$\,\% of the gas-phase inventory and does not affect the global carbon budget. Nevertheless, the spectral impact can be substantial: haze particles, even at sub-ppm mass fractions, nonlinearly enhance the atmospheric opacity through scattering and absorption, strongly modulating the observed transmission spectra.

\item All cases with $f_{\mathrm{UV}}=10$: a ten-fold stronger near-UV flux simultaneously restores vigorous CO$_2$ photolysis and produces abundant OH. The CO again rises with altitude, but the enhanced UV field partly photodissociates both CO and the newly formed CO$_2$, preventing the high-altitude build-up seen in the low-UV modern case.

\end{enumerate}

Hence, the CO decline in the modern, low-UV atmosphere is a natural consequence of UV shielding by O$_3$ plus efficient removal via $\mathrm{CO} + \mathrm{OH} \rightarrow \mathrm{CO}_2$, whereas all other simulations lack one (or both) of those conditions and therefore show a monotonically increasing CO with height.
We also note that there is no photochemical haze in the modern Earth-like scenarios. The high O$_2$ and O$_3$ levels maintain oxidizing conditions that suppress hydrocarbon polymerization, thereby precluding haze nucleation. Both HCAER and HCAER2\footnote{HCAER and HCAER2 are the two aerosol tracers used by the Atmos photochemistry code. They represent the direct condensation of the gas–phase precursors C$_4$H$_2$ and C$_5$H$_4$. Nonzero values indicate ongoing haze formation.} remain zero at all altitudes, even under enhanced UV flux ($f_{\mathrm{UV}} = 10$).
%, consistent with previous findings that haze formation is inhibited by atmospheric oxygen \citep[e.g.,][]{Arney.et.al.2016AsBio..16..873A}.}

Sulfur dioxide (SO$_2$) and hydrogen sulfide (H$_2$S) represent the most prevalent sulfur gases released during volcanic activity. The challenge of reconciling the observed sulfur mass-independent fractionation (S-MIF) in laboratory sulfur dioxide photochemistry with that recorded in the Archean rock record has persisted as a significant issue since its initial discovery \citep{Farquhar.et.al.2000Sci...289..756F,Farquhar.et.al.2001JGR...10632829F}. In particular, sulfur dioxide exhibits a complex UV absorption spectrum characterized by two prominent absorption features spanning 260 to 340\,nm and 165 to 235\,nm, which plays a crucial role in understanding atmospheric processes and climate dynamics.
The accumulation of SO$_2$ in the atmosphere of TOI-700\,d (see \autoref{fig:abundance_profiles}) is attributed to low UV emission from the host star. This is evident in the significantly higher concentration of sulfur dioxide when comparing scenarios with low ($f_\mathrm{UV}$=1) and high ($f_\mathrm{UV}$=10) UV flux. 
In O$_2$-rich atmospheres, absorption of ultraviolet photons by SO$_2$ can induce predissociation, whereby the excited molecule dissociates to SO + O*. This initiates the following chemical transformations:
\begin{equation}
    \begin{gathered}
        \text{SO}_2 + h\nu \; (\lambda < 220 \, \textrm{nm}) \xrightarrow\ \text{SO} + \text{O}^* \\
        \text{SO} + \text{O}_2 + \text{M} \xrightarrow\ \text{SO}_3 + \text{M}
    \end{gathered}
\label{eq:SO3}
\end{equation}

Although SO$_2$ absorbs UV radiation, albeit less efficiently compared to O$_3$, it can still accumulate in significant quantities. When combined with CH$_4$ and haze particles in the atmosphere, it may provide adequate protection against UV radiation, particularly in O$_2$-poor atmospheres.
%\cite{Kasting.et.al.1989OLEB...19...95K} suggests that the S$_8$ allotrope of sulfur could potentially serve as an additional shield against UV radiation in the Archean atmospheres, being formed as a by-product of a series of reactions during atmospheric SO$_2$ photolysis.

The production of SO$_3$ and its potential reaction with H$_2$O to form sulfuric acid (H$_2$SO$_4$) is a crucial process in planetary atmospheres. As highlighted by \cite{Meadows.et.al.2018}, while this reaction can proceed efficiently in Venus-like atmospheres, it is heavily dependent on the availability of water. In the absence of sufficient H$_2$O, the formation of H$_2$SO$_4$ aerosols is significantly limited. Moreover, the photochemical model used in this study does not simulate Venus-like atmospheres, particularly the formation of H$_2$SO$_4$ clouds and their feedback effects on atmospheric temperature structure.
Additionally, on Earth, the oxidation of SO$_2$ to SO$_3$ often serves as a bottleneck that controls the rate of sulfuric acid formation \citep{Lizzio.and.DeBarr.1997}. However, on planets orbiting M-dwarf stars, such as Proxima Centauri b, as studied by \cite{Meadows.et.al.2018}, the destruction of SO$_3$ may become the dominant process, potentially inhibiting the formation of H$_2$SO$_4$ clouds. This suggests that, while sulfuric acid clouds could still form under specific conditions, their formation may be less efficient or even absent in certain climate scenarios.

On present-day Earth, the ozone layer effectively absorbs radiation between 180 and 280 nm through the formation and destruction of O$_3$, as described in \autoref{eq:CH4}, and partially absorbs radiation between 280 and 310 nm.
This region, located in the stratosphere at approximately 10 to 50\,km above Earth's surface, acts as a natural shield against harmful UV radiation. 
This protection was likely essential for the early emergence of life on land, as it shielded organisms from harmful ultraviolet radiation \citep{Cockell.2000P&SS...48..203C}. However, alternative views suggest that such protection may not have been strictly required, since natural environments could also mitigate UV exposure. For instance, water bodies and geological cover can attenuate UV flux, particularly at the UV wavelengths, and may have provided effective refuges for early organisms or prebiotic molecules \citep{Estrela2018, Estrela2020, Ranjan.et.al.2022AsBio..22..242R}.

\begin{figure*}
\centering       
\includegraphics[width=0.99\textwidth]{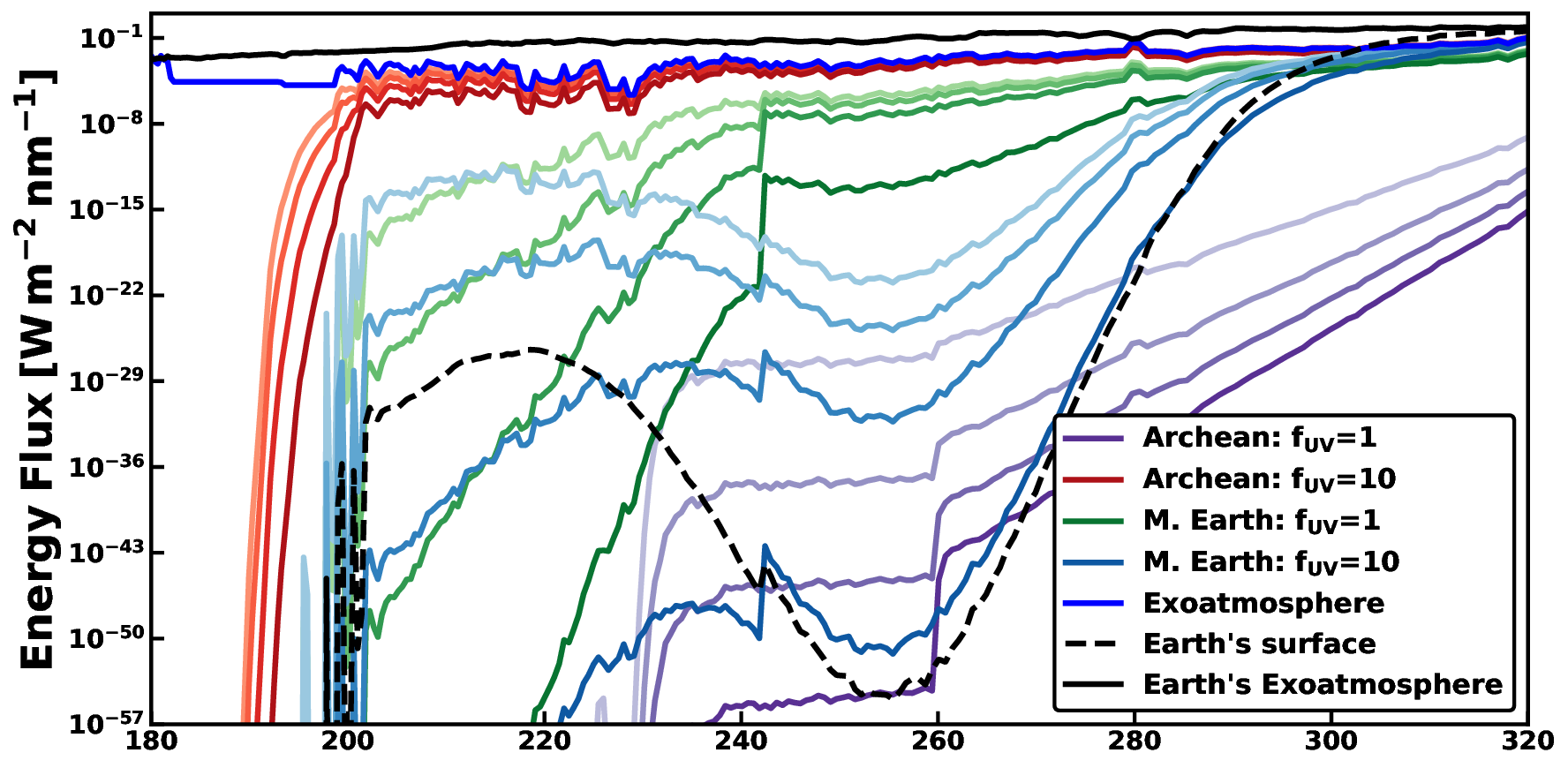}
\includegraphics[width=0.99\textwidth]{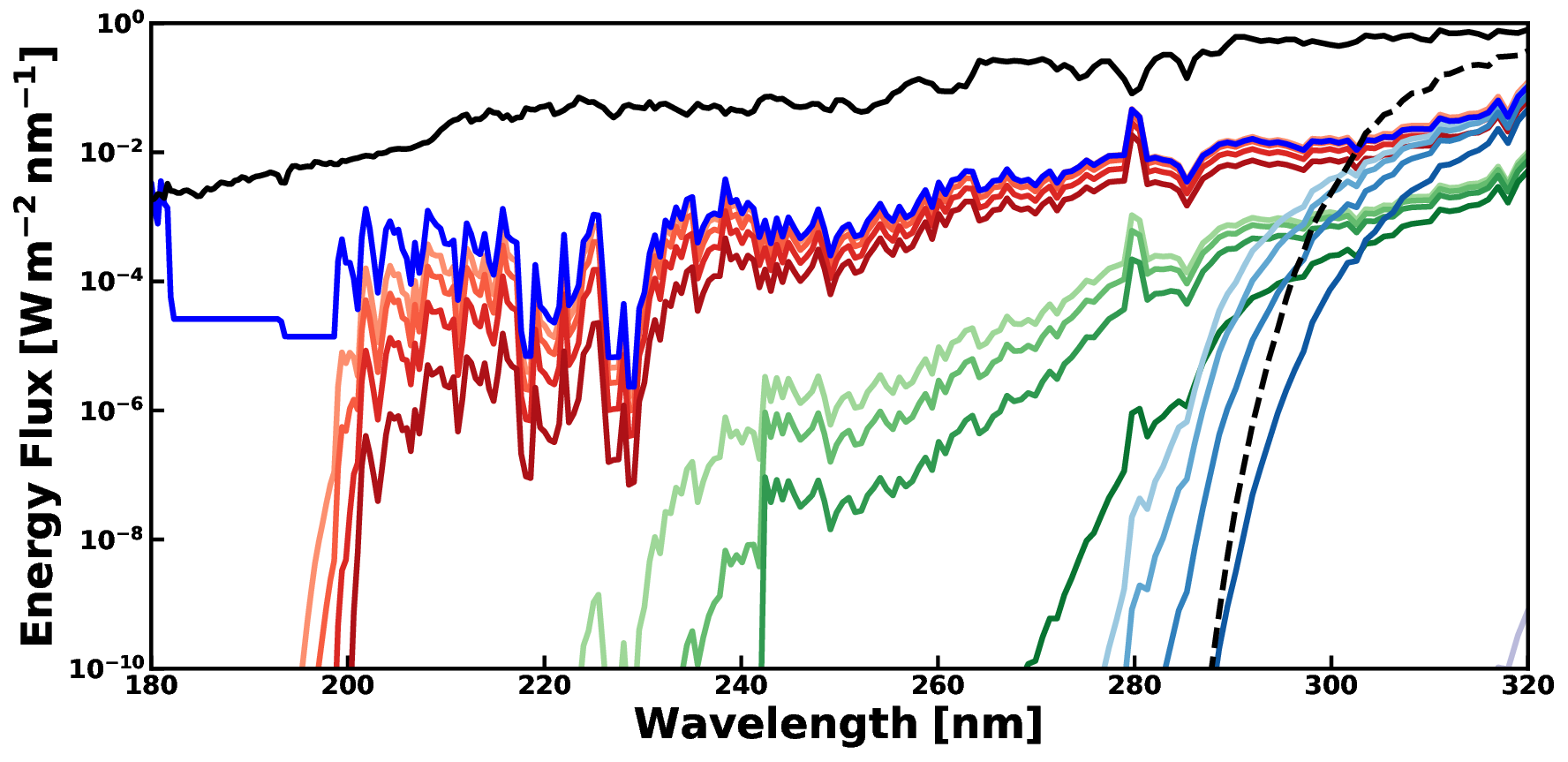}
\caption{Upper panel: UV flux reaching the surface of the planet TOI-700\,d for an Archean atmosphere, depicted in purple ($f_\mathrm{UV}$=1) and red ($f_\mathrm{UV}$=10), and a modern Earth-like atmosphere, in green ($f_\mathrm{UV}$=1) and blue ($f_\mathrm{UV}$=10). The different surface pressures, namely 4.0, 2.0, 1.0, and 0.5, are represented by tones gradually shifting towards lighter shades, indicating a transition from higher to lower pressure. The solar flux reaching both Earth's exosphere and surface is depicted by solid black line and dashed black line, respectively, primarily for the purpose of comparison. Lower panel: a blowup of the flux scale depicted on the upper panel.}
% The upper and lower panels can be distinguished solely by the energy flux scale displayed on the graphic.}
\label{fig:spectrum_UV}
\end{figure*}

A comparison of different atmospheric models reveals that only in the modern Earth-like atmosphere scenario with $f_\mathrm{UV}$=10, the ozone concentration is sufficient for the appearance of the Hartley band (200 and 320\,nm) (see \autoref{fig:spectrum_UV}). It was anticipated that ozone concentrations in the Archean model were expected to be extremely low, as well as in the modern Earth model with $f_\mathrm{UV}$=1.
This significant disparity in ozone concentration among models leads to variations in the UV radiation that penetrates the simulated atmospheres and reaches the planet's surface, potentially impacting the planet's habitability.

\begin{figure}
\centering       
\includegraphics[width=0.99\columnwidth]{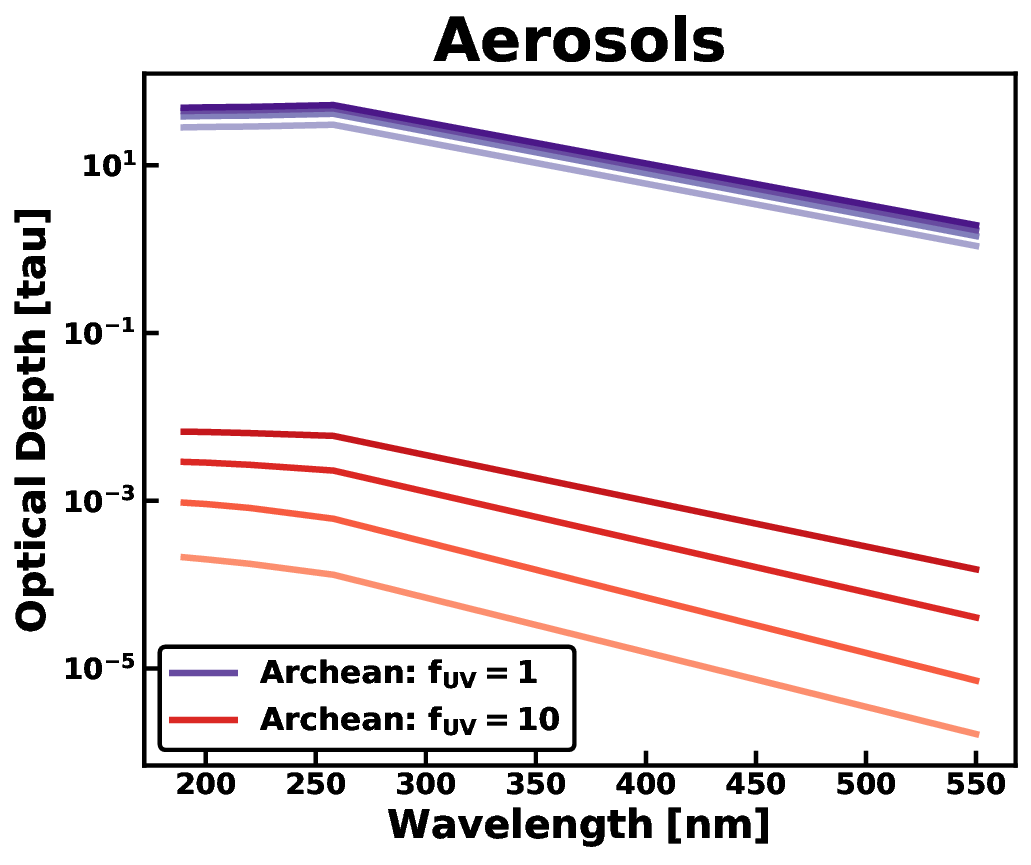}
\caption{Atmospheric optical depth as a function of wavelength for $f_\mathrm{UV}$=1 (purple curves) and $f_\mathrm{UV}$=10 (red curves) Archean atmospheres. Different surface pressures, namely 4.0, 2.0, 1.0, and 0.5, are represented by tones gradually shifting towards lighter shades, indicating a transition from higher to lower pressure. Under 10 times stellar UV, the aerosol column thins (lower $\tau$), consistent with non-monotonic haze production in CO$_2$-rich anoxic atmospheres where short-wave UV also enhances oxidative destruction of intermediates.}
\label{fig:tau_aerosols}
\end{figure}

In \autoref{fig:spectrum_UV}, we illustrate the incident radiation reaching the top of TOI-700\,d's atmosphere that penetrates down to its surface. Haze exhibits robust absorption, especially at blue and UV wavelengths, emphasizing its significant atmospheric role alongside ozone.
In the Archean model with low UV flux ($f_\mathrm{UV}$=1), the constituent molecules of haze play a more significant role than ozone photochemistry under high UV flux ($f_\mathrm{UV}$=10) conditions for a modern Earth-like atmosphere.  
Consequently, in the Archean model, the presence of a thick haze layer provides significant UV shielding, absorbing radiation at wavelengths below 230\,nm. This contrasts with the modern Earth scenario, where ozone primarily absorbs in the 200–310\,nm range, particularly through the Hartley band. As a result, in the low-UV regime, haze acts as the dominant protective mechanism against harmful radiation, reducing surface UV flux to levels lower than those on present-day Earth, especially under higher atmospheric pressure conditions.
Photolysis in the Archean model with $f_\mathrm{UV}$=10 leads to markedly lower concentrations of haze constituents than in the $f_\mathrm{UV}$=1 Archean model, resulting in reduced absorption of incident flux for wavelengths below 190\,nm.
This is evidenced by \autoref{fig:tau_aerosols}, which demonstrates a significant rise in haze concentration within TOI-700\,d's atmosphere under $f_\mathrm{UV}$=1 conditions compared to $f_\mathrm{UV}$=10. When assessing Archean models, the optical depth increases to 10$^4$.
Furthermore, in the Earth-like atmosphere model with $f_\mathrm{UV}$=1, there was minimal ozone production, failing to generate a discernible feature in the spectrum, such as the Hartley band. In this context, the surface UV in this case is lower only than in the Archean $f_{\mathrm{UV}}=10$ model.

Near-UV photons not only drive the production of haze precursors but also enhance their destruction via oxidizing pathways (e.g., O/OH from CO$_2$ photolysis), making net haze formation non-monotonic with incident UV radiation. Consistent with this mechanism, in our Archean simulations with enhanced stellar UV the organic haze column is reduced and the surface UV increases. While short-wavelength UV initiates methane photolysis, stronger UV in an anoxic, CO$_2$-bearing atmosphere also accelerates competing oxidative loss of key intermediates, lowering the aerosol yield. This non-linear behavior agrees with prior modeling of CO$_2$-rich Archean conditions \citep[e.g.,][]{Arney.et.al.2016AsBio..16..873A} and with broader evidence that the specific routes to particle formation remain uncertain and model-dependent \citep{Horst.2017JGRE..122..432H}.

\subsection{Biological effect} \label{sec:bio_effect}

\begin{figure*}
\centering       
\includegraphics[width=0.99\columnwidth]{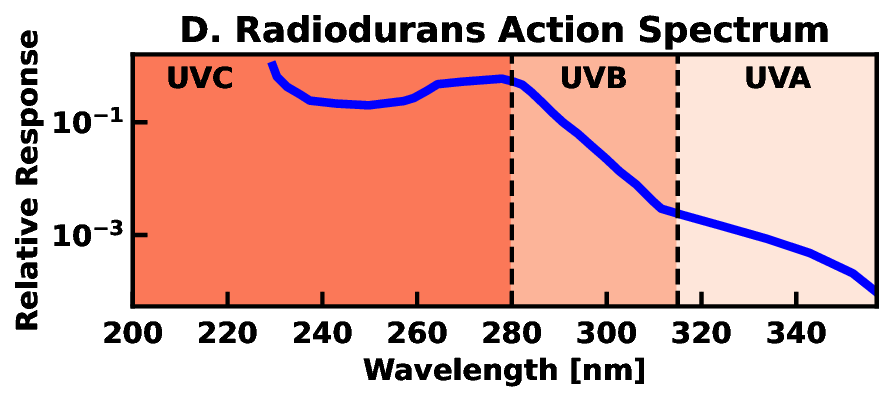}
\includegraphics[width=0.99\columnwidth]{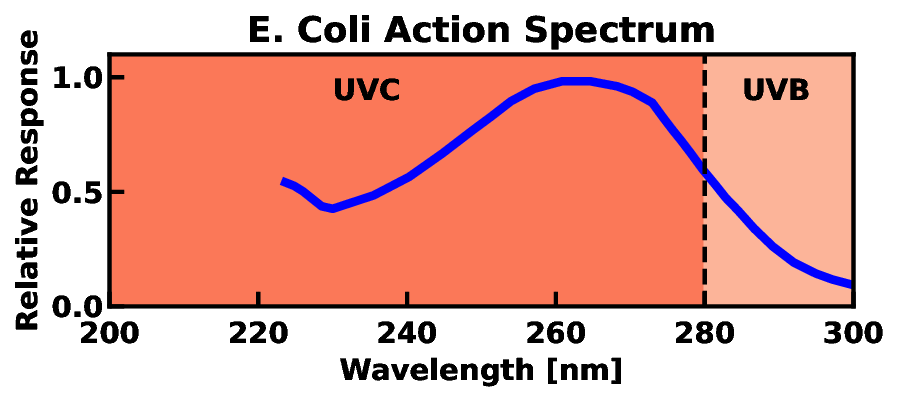}
\caption{Action spectra, or biological response, for \textit{D. radiodurans} (left) and \textit{E. coli} (right).}
\label{fig:Ecoli_Drdurans_action}
\end{figure*}

To assess the habitability potential of TOI-700\,d, we conducted an analysis of the UV radiation's impact on the planet's surface. We follow the  methodology previously utilized for the Kepler-96 \citep{Estrela2018} and Trappist-1 \citep{Estrela2020} systems. Our evaluation involves two key microorganisms: \textit{Deinococcus radiodurans}, known for its resilience to high UV doses, and \textit{Escherichia coli}, a widely researched bacterium. The biologically significant irradiance ($E_\text{eff}$), also referred to as fluence, is calculated based on the star's emission:
\begin{equation}
E_\text{eff} = \int_{\lambda_1}^{\lambda_2} F_\text{inc}(\lambda)\ S(\lambda)\ d\lambda \; .
    \label{eq:eff}
\end{equation}
\noindent In this equation $F_\text{inc}$ represents the total stellar incident flux at a specific wavelength $\lambda$, while $S(\lambda)$ is the action spectra of the bacteria (refer to \autoref{fig:Ecoli_Drdurans_action}).

To determine the bacteria's survival rate, we considered the total radiation dose accumulated by each bacterium during its generation time. We adopted a generation time of 20 minutes for \textit{E. coli} \citep{Arp1980} and 100 minutes for \textit{D. radiodurans} \citep{Mattimore1995}.

The survival rate for 37\% and 10\% of the bacterial population as a function of the UV radiation received is given by \cite{Gascon1995}. Specifically, for \textit{D. radiodurans}, these rates correspond to doses of 338 and 553\,J\,m$^{-2}$, while for \textit{E. coli}, the rates are 17.3 and 22.6\,J\,m$^{-2}$, respectively. The UV wavelength range considered by \cite{Gascon1995} spans from 230 to 320\,nm for \textit{D. radiodurans} and from 224 to 300\,nm for \textit{E. coli}. These wavelength bounds were determined based on the bacteria's action spectrum and the absorption cross-section of the molecules under study.

To estimate the potential habitability of TOI-700\,d in the given scenarios, we calculated the survival rates of the two bacteria on the planet's surface. Applying \autoref{eq:eff}, we determined the radiation doses that the bacteria would experience over a generation period (20 minutes for \textit{E. coli} and 100 minutes for \textit{D. radiodurans}). 

\begin{figure*}
    \centering         \includegraphics[width=0.99\textwidth]{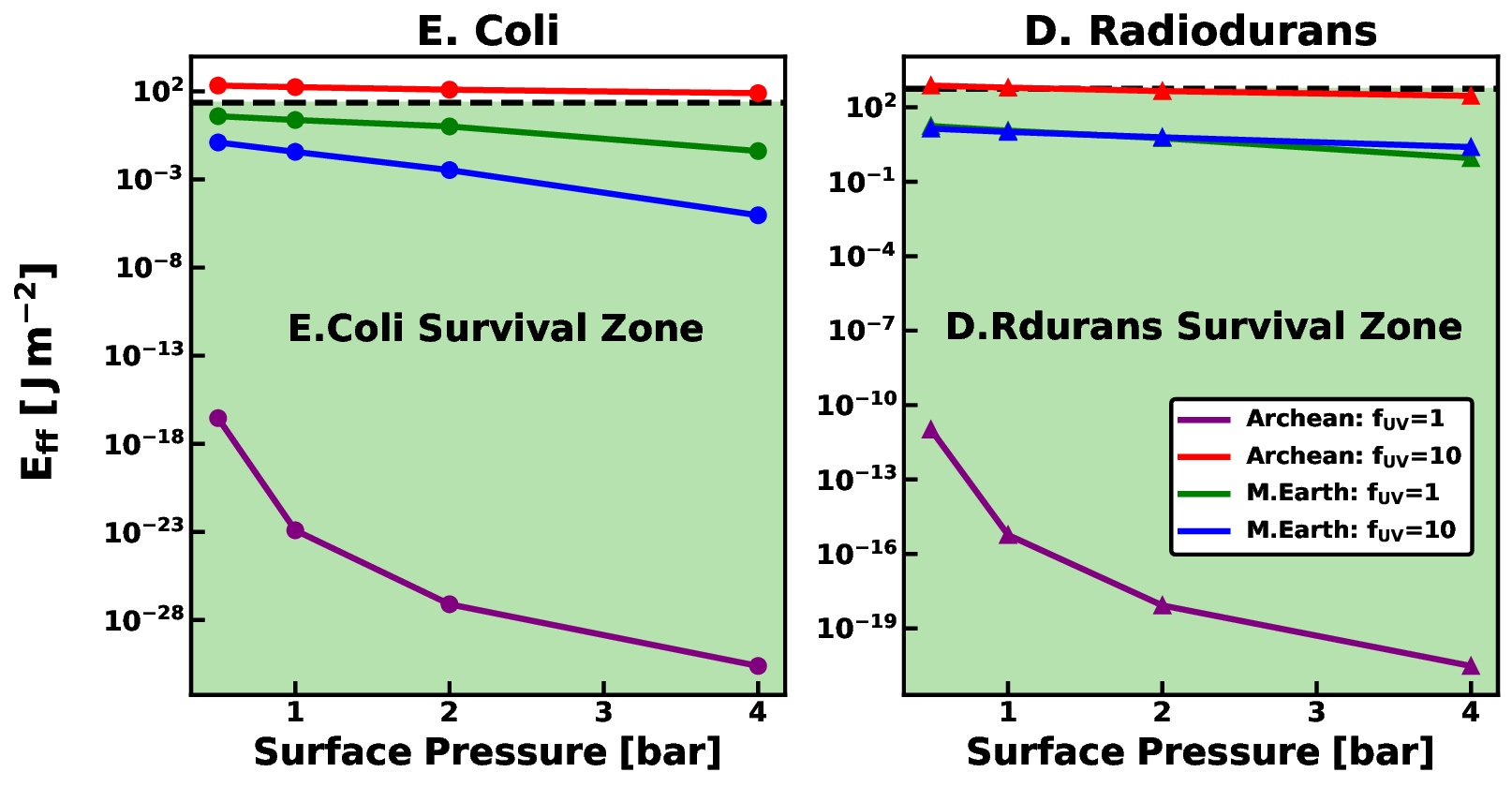}
    \caption{Biologically Effective Irradiance as a function of surface pressure based on results of \autoref{tab:eff}. The $E_\mathrm{ff}$ is calculated over a generation period for the two types of bacteria. The green area delineates the conditions where 10\% of \textit{E. coli} (left panel) and \textit{D. radiodurans} (right panel) would thrive, whereas the white region represents an environment unsuitable for bacterial survival.}
    \label{fig:Ecoli_Drdurans_survival}
\end{figure*}

The detailed results are listed in \autoref{tab:eff}, while \autoref{fig:Ecoli_Drdurans_survival} displays the derived values with the boundaries delineating the survival rates specifically for 10\% of the bacterial population. Concerning TOI-700\,d, among the four modeled atmospheres with different pressures, only the Archean atmosphere with $f_\text{UV}$=10 fails to sustain the survival of \textit{E. coli} (across all pressures) and \textit{D. radiodurans} (at pressures of 0.5 and 1.0 bar) under intense UV exposure.

\begin{table}
\caption{Biologically effective irradiance in [J\,m$^{-2}$] for TOI-700\,d atmospheric models.}
\begin{center}
    \begin{tabular}{c|lcc}
    \hline
    \multicolumn{4}{c}{UV=1}\\
    \hline
    Pressure  & Bacterium & Archean & Modern Earth \\
    
    [bar] & & [J\,m$^{-2}$] & [J\,m$^{-2}$]\\
    \hline
    0.5 &	\textit{E. coli} & 2.86$\times10^{-14}$ & 3.88 \\
     & \textit{D. radiodurans} & 1.03$\times10^{-11}$ & 17.67 \\
    1.0 &	\textit{E. coli} & 1.24$\times10^{-23}$ & 2.35 \\
    & \textit{D. radiodurans} & 5.82$\times10^{-16}$ & 11.45 \\
    2.0 &	\textit{E. coli} & 7.89$\times10^{-28}$ & 1.02 \\
    & \textit{D. radiodurans} & 8.41$\times10^{-16}$ & 5.67\\
    4.0 &	\textit{E. coli} & 2.47$\times10^{-31}$ & 0.050 \\
    & \textit{D. radiodurans}	& 3.05$\times10^{-21}$	& 0.89\\
    \hline
    \multicolumn{4}{c}{UV=10}\\
    \hline
    Pressure  & Bacterium & Archean & Modern Earth \\
        
    [bar] & & [J\,m$^{-2}$] & [J\,m$^{-2}$]\\
    \hline
    0.5 &	\textit{E. coli} & 212.33 & 0.12 \\
     & \textit{D. radiodurans} & 752.32 & 13.89\\
    1.0 &	\textit{E. coli} & 173.10 & 0.037 \\
    & \textit{D. radiodurans} & 620.60 & 10.20\\
    2.0 &	\textit{E. coli} & 124.01 & 0.0033 \\
    & \textit{D. radiodurans} & 450.60 & 6.17\\
    4.0 &	\textit{E. coli} & 78.53 & 9.27$\times10^{-6}$\\
    & \textit{D. radiodurans} & 288.82 & 2.50\\
    \hline
    \end{tabular}
\raggedright\textbf{Note:} the $E_\mathrm{ff}$ is calculated over a generation period for both types of bacteria.
\end{center}
\label{tab:eff}
\end{table}

As highlighted by \cite{Estrela2020}, TRAPPIST-1, a star smaller and cooler than TOI-700 but highly active with constant flare activity, requires an ozone layer for the survival of \textit{E. coli} bacteria on habitable planets within the system. The existence of this layer is a critical aspect for TOI-700\,d to maintain an Earth-like atmosphere. The presence and strength of UV radiation are key contributors to ozone formation in the atmosphere, as discussed in Subsection~\ref{sec:photoclima}. This is achieved only when $f_\text{UV}$=10. In this scenario, the microorganisms could thrive in such atmosphere.

As discussed previously, denser hazes avert catastrophic cooling of the planet. \cite{Arney.et.al.2016AsBio..16..873A} concluded that hazes could potentially enhance planetary habitability by providing UV protection, reducing surface UV flux by approximately 97\% compared to a planet devoid of haze. This reduction could potentially sustain terrestrial organisms that existed 2.7-2.6 billion years ago. In our analysis, the haze in $f_\text{UV}$=1 Archean atmosphere lowers the Biologically Effective Irradiance to $1.24\times$ 10$^{-23}$\,J\,m$^{-2}$ for \textit{E. coli} and to $5.82\times$ 10$^{-16}$\,J\,m$^{-2}$ for \textit{D. radiodurans}, at a pressure of 1\,bar. No other scenario produced such low values, indicating that haze, in this context, offers a significantly greater protective barrier than ozone. However, in scenarios where haze and ozone are less prevalent, such as in the Archean atmosphere model with $f_\text{UV}$=10, only the bacterium \textit{D. radiodurans} could survive with atmospheres under pressures of 2.0 and 4.0\,bar.

\subsection{Simulated Transmission Spectra}
\label{sec:transmission_spectra}

\begin{figure*}
\centering       
\includegraphics[width=0.99\textwidth]{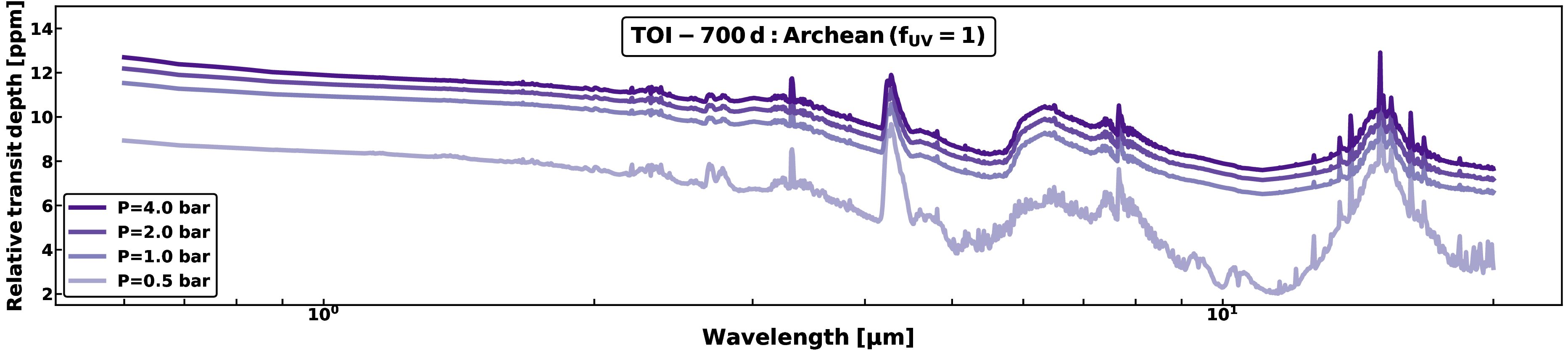}    \includegraphics[width=0.99\textwidth]{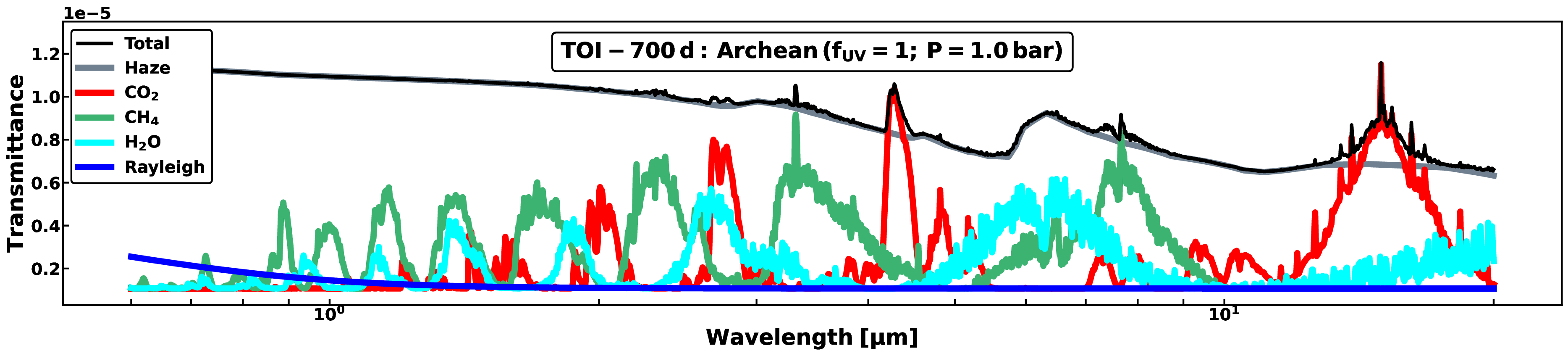}
\includegraphics[width=0.99\textwidth]{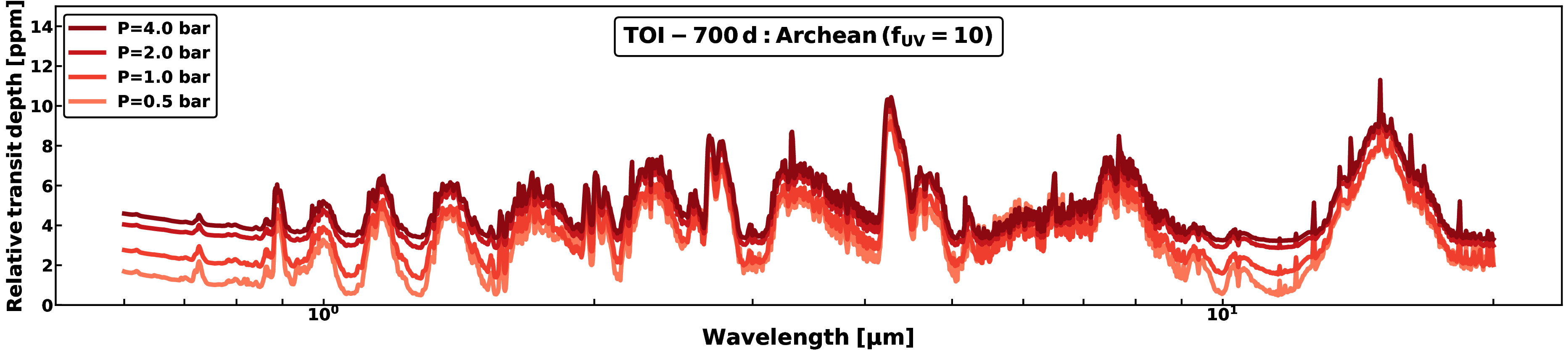}
\includegraphics[width=0.99\textwidth]{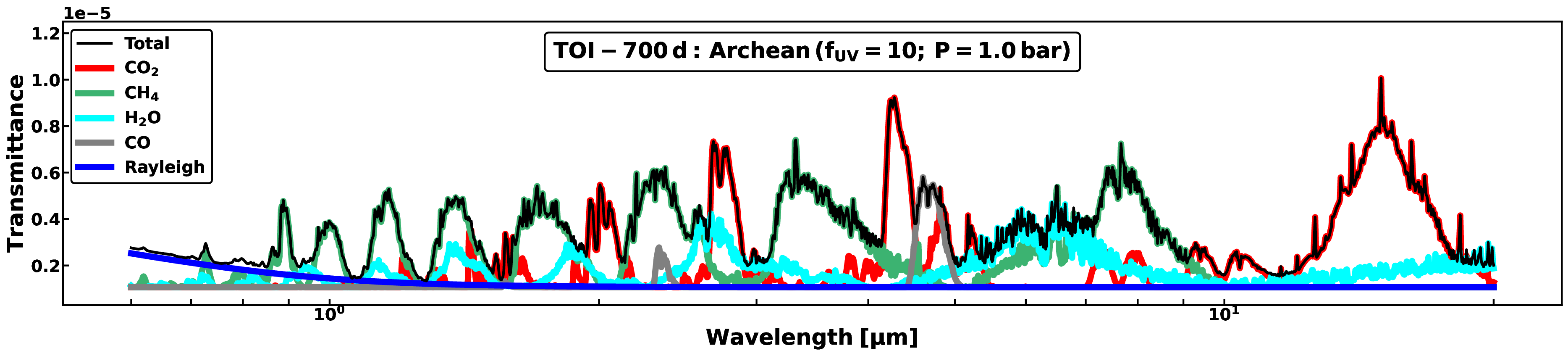}
\caption{TOI-700\,d Archean atmosphere's transmission spectra (top first and third panels) and transmittance (second and fourth panels) for the most abundant molecules. The Archean atmospheres with $f_\mathrm{UV}$=1 and $f_\mathrm{UV}$=10 are depicted in purple (first panel) and red (third panel), respectively. The different surface pressures, i.e., 4.0, 2.0, 1.0, and 0.5\,bar, are represented by tones gradually shifting towards lighter shades, indicating a transition from higher to lower pressure.
In panels 2 and 4, in addition to the most prevalent molecules, the influences of haze and Rayleigh scattering are also visible. The combined contribution is depicted by the black line.}
\label{fig:spectra_archean}
\end{figure*}

\begin{figure*}
\centering       
\includegraphics[width=0.99\textwidth]{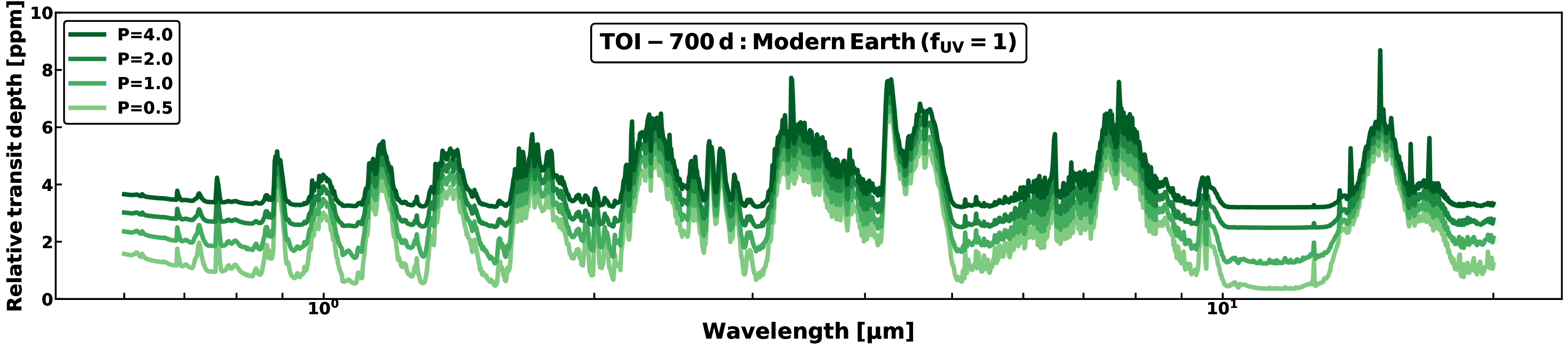}
\includegraphics[width=0.99\textwidth]{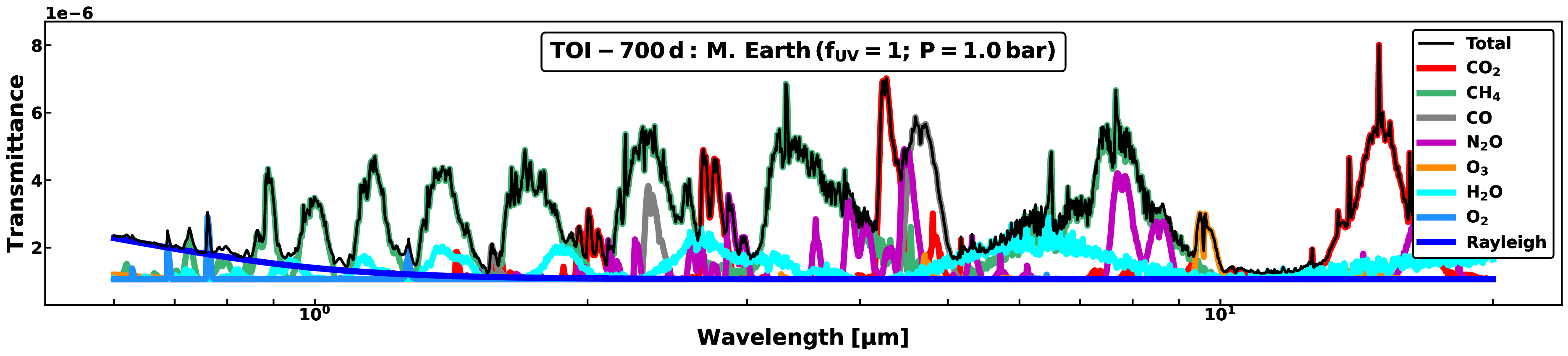}
\includegraphics[width=0.99\textwidth]{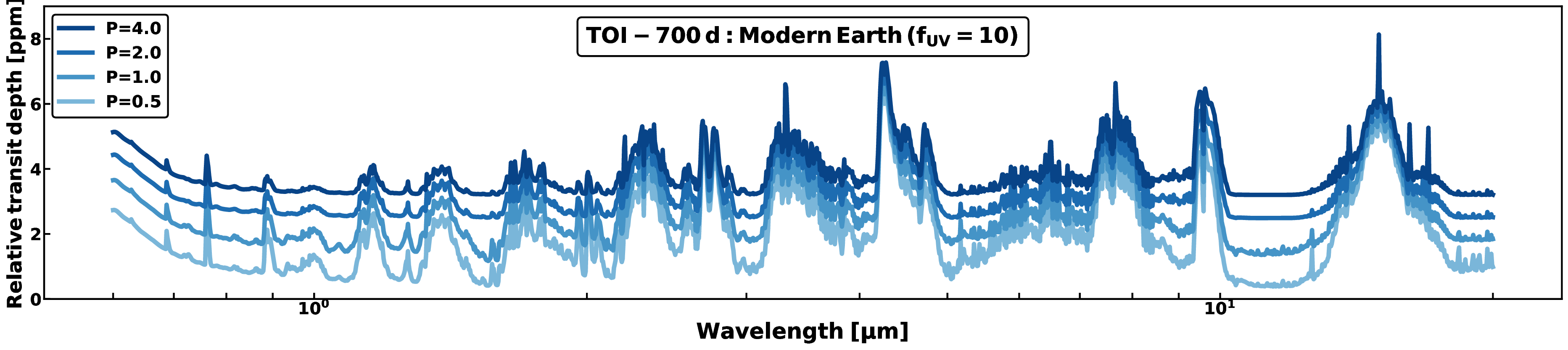}
\includegraphics[width=0.99\textwidth]{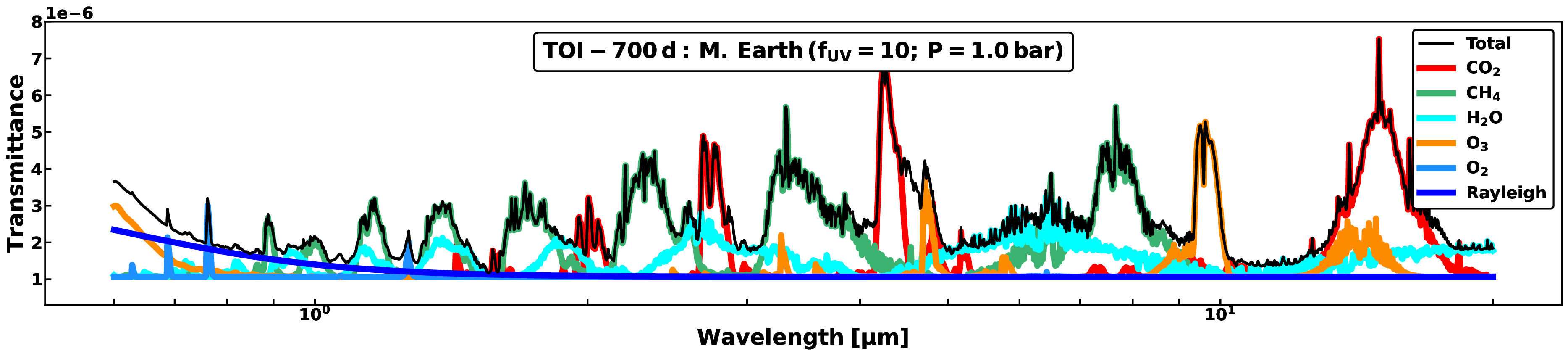}
\caption{Same as \autoref{fig:spectra_archean} for the modern Earth-like atmosphere model of TOI-700\,d.
%The first and third panels show the transmission spectra, whereas the second and fourth panels display the transmittance for the most prevalent molecules. The Earth-like atmospheres with $f_\mathrm{UV}$=1 and $f_\mathrm{UV}$=35 are depicted in green and blue, respectively. The surface different pressures, i.e., 4.0, 2.0, 1.0, and 0.5\,bar, are represented by tones gradually shifting towards lighter shades, indicating a transition from higher to lower pressure.
}
\label{fig:spectra_m.earth}
\end{figure*}

In Figures \ref{fig:spectra_archean} and \ref{fig:spectra_m.earth}, we present the transmission spectra from the atmospheric models, including the corresponding spectrum for each scenario featuring the most significant spectral lines at a pressure of 1\,bar. In addition to displaying the spectral lines of chemical compositions, these figures also illustrate the Rayleigh scattering slope. 
\cite{Benneke.and.Seager.2012ApJ...753..100B} proposed using the Rayleigh scattering slope as a means to constrain atmospheric pressure on exoplanets. However, according to \cite{Arney.et.al.2016AsBio..16..873A}, this approach may not be feasible for planets with hydrocarbon hazes due to the strong absorption effects at short wavelengths caused by these hazes. The hazes exhibit a significant spectral impact at shorter wavelengths, primarily due to their pronounced blue and UV absorption. The presence of thick hazes notably diminishes the strength of gaseous absorption features in transit transmission spectra, especially at shorter wavelengths where these hazes are optically thick. However, while hazes obscure key biosignature lines, they also enhance the overall detectability of the atmosphere by increasing scattering and transit depth, making the presence of an atmosphere more apparent. 
This characteristic is observed in the $f_\mathrm{UV}$=1 Archean atmosphere model. 

The Rayleigh slope in transmission spectra arises from the stronger scattering of shorter-wavelength light by particles much smaller than the wavelength itself. Consequently, blue and violet wavelengths experience more scattering than red wavelengths. In cases with high ozone concentrations, such as the modern Earth-like atmosphere model at $f_\mathrm{UV}=10$, the presence of O$_3$ significantly steepens the Rayleigh slope. This effect is further evidenced by the appearance of distinct absorption lines (e.g., near 10\,$\mu$m) in the model subjected to intense UV flux; by contrast, at $f_\mathrm{UV}=1$, these lines are weaker and virtually absent for Archean atmospheres.

Moreover, transmission spectra are strongly impacted by surface pressure. In higher-pressure atmospheres, increased pressure broadening leads to a reduction in scale height, resulting in weaker spectral features. By contrast, lower-pressure atmospheres exhibit larger scale heights, extending the atmospheric column and producing stronger absorption signals. This pressure-dependent effect is clearly visible in Figures \ref{fig:spectra_archean} and \ref{fig:spectra_m.earth}, where higher-pressure cases lead to more muted spectral lines, whereas lower-pressure cases enhance their amplitudes.

To distinguish planets with compositions resembling modern Earth-like or Archean atmospheres, the identification of CH$_4$ features could prove to be a crucial observational constraint. 
The strength of methane features correlates with its abundance in the atmosphere, as demonstrated by the comparison of feature heights. However, the presence of CH$_4$ will depend predominantly on the received UV flux rather than the type of atmosphere modeled in this study. Conversely, although CO$_2$ features are prominent in all simulated scenarios, irrespective of their abundance, relying solely on CO$_2$ detection to characterize the type of atmosphere is inadequate, as the presence of CO$_2$ is conspicuous in all of the simulated cases. This observation was also noted by \cite{Suissa2020b} in their simulations of ``Early Mars'' atmosphere.

From the particular scenarios investigated in our study, we can deduce that CO$_2$ is less sensitive to fluctuations in temperature and pressure. Consequently, its absorption characteristics remain relatively stable regardless of atmospheric conditions on the planet, which stands in contrast to the variability seen in methane. However, detecting absorption lines of CO$_2$ and CH$_4$ can pose difficulties. In the case of a moist stratosphere, the overlap of H$_2$O spectral bands with those of CO$_2$ and CH$_4$ requires an extended observation period to reach a desired level of detection significance \citep[see][]{Mikal-Evans.et.al.2022MNRAS.510..980M}.

The simultaneous detection of CH$_4$ and CO$_2$ is particularly relevant to evaluate biosignatures under anoxic conditions akin to Archean Earth. Although our models assume the coexistence of these gases, their observation in real exoplanetary atmospheres -- particularly when found in disequilibrium or with a notable absence of CO -- could strongly indicate biological activity. This is because non-biological sources of methane typically produce carbon monoxide as a by-product. Thus, atmospheres abundant in CO$_2$ and CH$_4$ but scarce in CO remain significant indicators of biotic processes, underscoring the importance of monitoring these gases for potential extraterrestrial life \citep[see][]{Krissansen-Totton.et.al.2018SciA....4.5747K, Mikal-Evans.et.al.2022MNRAS.510..980M, Rotman.et.al.2023ApJ...942L...4R}."

Consequently, our findings align with \cite{Suissa2020b} regarding spectral line contrasts peaking at less than 15\,ppm. Thus, characterizing TOI-700\,d with JWST appears unfeasible due to its instrumental noise ranging from 10-20 parts per million at 1$\sigma$ \citep{Greene.et.al.2016ApJ...817...17G}. According to estimates of \cite{Suissa2020b}, hundreds of JWST transits or eclipses would be required, exceeding the nominal lifetime of the observatory. Therefore, characterizing TOI-700\,d's atmosphere remains an extremely challenging task even for next-generation telescopes.

\section{Summary and conclusions} 
\label{sec:conclusion}

In this work, we coupled the 1D Photochemical Model with the Atmos Climate Model to investigate the atmospheric conditions of TOI-700\,d under two distinct scenarios: an analogue of the Archean Earth and a modern Earth-like atmosphere. Although both scenarios correspond to biologically active periods in Earth's history, our goal is to assess whether such environments could remain habitable under the photochemical UV conditions of an M dwarf. To do this, we explore different levels of stellar UV flux -- quiescent ($f_{\mathrm{UV}}=1$) and enhanced flare ($f_{\mathrm{UV}}=10$) conditions -- across a range of atmospheric pressures (0.5, 1.0, 2.0, and 4.0 bar). Our key findings are summarized as follows:

\begin{itemize}

\item \textbf{Atmospheric Composition and UV Shielding:} Archean (methane-rich) and modern Earth-like atmospheres exhibit diverse photochemical pathways, significantly influenced by the incoming UV flux. In the quiescent-state Archean scenario, high haze production was found to provide efficient shielding against harmful UV radiation, whereas modern Earth-like atmospheres under high UV flux relied more on ozone formation to attenuate surface-level UV.
\item \textbf{Surface Conditions and Potential for Life:} the temperature profiles suggest that most simulated atmospheres could sustain surface temperatures compatible with liquid water, an essential requirement for life. Through biologically effective irradiance calculations, we determined that the survival of model bacteria such as \textit{Escherichia coli} and \textit{Deinococcus radiodurans} is strongly dependent on a combination of atmospheric pressure, haze abundance, and ozone column density. Specifically, thick haze layers (in the context of an Archean atmosphere without intensified UV exposure) or robust ozone formation (in a modern Earth-like environment with high UV radiation) can provide significant protection against surface-level UV exposure.
\item \textbf{Transmission Spectra and Biosignature Detection:} we used the Planetary Spectrum Generator (PSG) to produce synthetic transmission spectra for each model. Haze particles can significantly obscure spectral features at shorter wavelengths, potentially masking key biosignatures such as CH$_4$ and O$_2$, while ozone absorption bands become prominent only when UV flux levels are sufficient in an Earth-like atmosphere. Our results indicate that the detectability of biosignatures on TOI-700\,d could be extremely challenging with current or near-future instrumentation, as the strongest spectral lines remain at the level of a few to a few tens of parts per million. 
Furthermore, our transmission spectrum analysis underscores the significant role of surface pressure in shaping observed spectral features: higher pressures reduce scale height and thus diminish absorption line depths, whereas lower pressures yield more pronounced spectral features due to an extended atmospheric column. These factors -- haze scattering, ozone absorption, and variations in stellar UV flux -- interact in a complex manner, posing significant challenges for the detection of biosignatures in M-dwarf planetary systems.
\item \textbf{Implications for General Exoplanetary Systems:} while TOI-700\,d is a particularly compelling target due to its relatively inactive host star and Earth-like insolation, our study offers broader insights for exoplanets orbiting M dwarfs. Dense hazes and high levels of ozone can each play pivotal roles in rendering a planetary surface habitable, yet they also complicate detection of biosignatures. For many M-dwarf systems, flare events -- and the associated UV flux boosts -- could lead to rapid shifts in atmospheric chemistry, either destroying protective layers or driving the formation of new ones. These processes are critical to consider when searching for signs of life elsewhere in the universe.

\end{itemize}

Taken together, our results highlight the interconnected effects of photochemistry, atmospheric pressure, haze formation, and UV flux on the habitability and spectroscopic detectability of exoplanetary atmospheres. Although TOI-700\,d exemplifies these complexities under relatively benign stellar conditions, analogous mechanisms likely operate in numerous M-dwarf systems. Ongoing and future multi-wavelength observational campaigns -- aimed at capturing transits, flare events, and high-resolution spectral observations -- will be essential to refine these models and advance our understanding of habitable environments beyond the Solar System.

\begin{acknowledgments}
\small
The authors gratefully acknowledge Dr. Paul B. Rimmer for his thoughtful suggestions, which helped improve the manuscript.
We also thank Abel Granjeiro de Souza for the work on TOI-700\,d of his undergraduate thesis, which set the stage for the investigations undertaken here. 
VYDS and AV acknowledge the partial financial support received from Brazilian FAPESP grants \#2021/14897-9, \#2018/04055-8 \#2021/02120-0, and \#2024/03652-3, as well as CAPES and MackPesquisa funding agencies. R.E. research was carried out at the Jet Propulsion Laboratory, California Institute of Technology, under a contract with the National Aeronautics and Space Administration (80NM0018D0004).
\end{acknowledgments}


\appendix

\autoref{tab:inital_produced_mix_ratios} shows the initial conditions and produced surface mixing ratios of the nine major long-lived species for the specific case where the pressure is 1 bar. All input and output data from this study can be found at \href{https://doi.org/10.5281/zenodo.13947863}{10.5281/zenodo.13947863}.

\begin{table}[h]
\centering
\caption{Initial conditions and produced surface mixing ratios of long-lived species for the specific case where the pressure is 1\,bar, with emphasis on the main species.}
\begin{tabular}{c|c|c|c|c|c|c}
\hline
\multicolumn{1}{c|}{} & \multicolumn{2}{c|}{\textbf{Initial conditions}} & \multicolumn{4}{c}{\textbf{Produced surface mixing ratios}} \\
\textbf{Species} & \multicolumn{2}{c|}{\textbf{Surface mixing ratios}} & \multicolumn{4}{c}{\textbf{($P=1\,\mathrm{bar}$)}} \\
%\hline  % Retiramos o \hline aqui para não inserir a linha sob "Species"
\cline{2-7}  % Linhas apenas das colunas 2 até 7
 & \textbf{Archean} & \textbf{Modern Earth} & \textbf{Archean} & \textbf{Modern Earth} & \textbf{Archean} & \textbf{Modern Earth} \\
 & & & ($f_\mathrm{uv}=1$) & ($f_\mathrm{uv}=1$) & ($f_\mathrm{uv}=10$) & ($f_\mathrm{uv}=10$) \\
\cline{1-7} % Mantém as linhas intactas nas outras colunas
O$_2$ & 1.00$\times10^{-8}$ & 2.10$\times10^{-1}$ & 1.00$\times10^{-8}$ & 2.10$\times10^{-1}$ & 1.00$\times10^{-8}$ & 2.10$\times10^{-1}$ \\ \cline{1-7}
O$_3$ & 5.94$\times10^{-12}$ & 3.01$\times10^{-8}$ & 2.17$\times10^{-20}$ & 1.05$\times10^{-8}$ & 3.96$\times10^{-12}$ & 3.89$\times10^{-8}$ \\ \cline{1-7}
CO & 1.20$\times10^{-4}$ & 2.35$\times10^{-7}$ & 1.20$\times10^{-4}$ & 5.57$\times10^{-3}$ & 1.20$\times10^{-4}$ & 2.97$\times10^{-5}$ \\ \cline{1-7}
CO$_2$ & 2.00$\times10^{-2}$ & 3.85$\times10^{-4}$ & 2.00$\times10^{-2}$ & 3.60$\times10^{-4}$ & 2.00$\times10^{-2}$ & 3.76$\times10^{-4}$ \\ \cline{1-7}
H$_2$O & 1.17$\times10^{-2}$ & 7.05$\times10^{-3}$ & 1.17$\times10^{-2}$ & 7.05$\times10^{-3}$ & 6.18$\times10^{-3}$ & 7.05$\times10^{-3}$ \\ \cline{1-7}
CH$_4$ & 2.00$\times10^{-3}$ & 1.79$\times10^{-6}$ & 2.00$\times10^{-3}$ & 2.58$\times10^{-3}$ & 2.00$\times10^{-3}$ & 2.12$\times10^{-4}$ \\ \cline{1-7}
SO$_2$ & 2.22$\times10^{-10}$ & 1.82$\times10^{-10}$ & 2.51$\times10^{-10}$ & 2.35$\times10^{-10}$ & 3.02$\times10^{-10}$ & 2.57$\times10^{-10}$ \\ \cline{1-7}
SO$_3$ & 2.74$\times10^{-19}$ & 2.49$\times10^{-19}$ & 1.75$\times10^{-25}$ & 2.84$\times10^{-20}$ & 1.503$\times10^{-19}$ & 2.96$\times10^{-20}$ \\ \cline{1-7}
N$_2$O & --- & 3.39$\times10^{-7}$ & --- & 9.01$\times10^{-6}$ & --- & 1.73$\times10^{-6}$ \\ \cline{1-7}
\end{tabular}
\label{tab:inital_produced_mix_ratios}
\end{table}

\bibliography{main}{}
\bibliographystyle{aasjournal}



\end{document}